%% file: main.tex
\documentclass[twocolumn]{autart}
\usepackage[utf8]{inputenc}
\usepackage{graphicx}    
\usepackage{amsmath, amsfonts}
\usepackage{mathtools}
\usepackage{accents}
\usepackage{url}
\usepackage{tikz,pgfplots}
\usepackage{amsmath}
\usetikzlibrary{arrows}

\usepackage[textwidth=1.5in]{todonotes}
\DeclareMathOperator*{\argmin}{arg\,min}
\DeclarePairedDelimiter{\ceil}{\lceil}{\rceil}
\DeclarePairedDelimiter{\floor}{\lfloor}{\rfloor}

\newtheorem{assump}{Assumption}
\newtheorem{probstatement}{Problem Statement}
\newtheorem{remark}{Remark}
\newtheorem{example}{Example}

\definecolor{mycolor1}{RGB}{230,97,1}
\definecolor{mycolor2}{RGB}{178,171,210}
\definecolor{mycolor3}{RGB}{253,184,99}
\definecolor{mycolor4}{RGB}{94,60,153}

\renewcommand{\P}{\mathbb{P}}
\newcommand{\E}{\mathbb{E}}
\newcommand{\Rho}{\mathrm{P}}
\newcommand{\ubar}[1]{\underaccent{\bar}{#1}}

\newcommand{\defineas}{\coloneqq}

\newcommand{\vdiff}[3]{\Delta^{#1}_{#2}{#3}}

\mathtoolsset{showonlyrefs}

\begin{document}
\begin{frontmatter}
\title{Resource Aware Pricing for Electric Vehicle Charging}
\vspace{-.5cm}
\author{Cesar Santoyo}, \author{Gustav Nilsson}, and \author{Samuel Coogan}

\thanks{The authors are with the School of Electrical \& Computer Engineering, Georgia Institute of Technology, Atlanta, GA, 30318, USA. S. Coogan is also with the School of Civil and Environmental Engineering, Georgia Institute of Technology, Atlanta, GA, 30318, USA. {\tt \{csantoyo, gustav.nilsson, sam.coogan\}@gatech.edu}. This work was partially supported by NSF under Grant \#1931980. C. Santoyo was supported by the NSF Graduate Research Fellowship Program under Grant No. DGE-1650044.}
\thanks{A preliminary version of this paper was in part published in the proceedings of the IFAC World Congress 2020~\cite{santoyo2020multi}.}

\begin{abstract}
Electric vehicle charging facilities offer electric charge and parking to users for a fee. Both parking availability and electric charge capacity are constrained resources, and as the demand for charging facilities grows with increasing electric vehicle adoption, so too does the potential for exceeding these resource limitations. In this paper, we study how prices set by the charging facility impact the likelihood that resource constraints are exceeded. Specifically, we present probabilistic bounds on the number of charging spots and the total power supply needed at a facility based on the characteristics of the arriving vehicles. We assume the charging facility either offers a set of distinct and fixed charging rates to each user or allows the user to decide a charging deadline, from which a charging rate is determined.  
Users arrive randomly, requiring a random amount of charge. Additionally, each user has a random impatience factor that quantifies their value of time, and a random desired time to stay at a particular location. 
Assuming rational user behavior, and with knowledge of the probability distribution of the random parameters, we present high-confidence bounds on the total number of vehicles parked at the station and the aggregate power use of all vehicles actively charging. We demonstrate how these bounds can be used by a charging facility to determine appropriate prices  and investigate through a Monte-Carlo simulation case study the tightness of the bounds.
\end{abstract} 

\end{frontmatter}
\maketitle

\section{Introduction}
The electric vehicle (EV) revolution promises to transform transportation and mobility. This has been catalyzed by improved affordability of electric vehicles (EVs) such that \cite{bloombergnef} predicts that by 2040 the global new vehicle sales will be comprised of 58\% EVs and the global passenger vehicle market will be 31\% electric.
With the growing numbers of electric vehicles, the demands on charging facilities will be greater. 

The EV charging problem can be classified into the following categories: EV usage in the context of smart grid or vehicle-to-grid, EV charging network design, EV charging facility pricing, and EV routing \cite{sassi2017electric},\cite{alizadeh2018retail}. Note that while these classifications are useful, there exists literature which addresses more than one of these categories at once. Furthermore, the EV charging problem has been addressed within game-theoretic, optimization, and control system frameworks while adhering to capacity constraints, e.g., in \cite{ortega2012electric}, \cite{wu2011vehicle}, \cite{zhang2015game}. 
For example, in \cite{li2013distribution}, charging management is performed by solving a social welfare nonlinear optimization problem subject to power constraints. Here, the distribution locational marginal pricing provides an effective way for mitigating charging facility congestion. In \cite{gan2012optimal}, scheduling electric vehicle charging is formulated as an optimal control problem which algorithmically converges to optimal charging profiles which are cognizant of power constraints. The paper~\cite{bae2011spatial} considers a spatiotemporal model for rapid charging facilities. There, the authors propose a queuing theoretic model which predicts the demand on charging facilities using the fluid traffic model such that the arrival rate of users is not known a priori. In~\cite{moradipari2019pricing}, the authors study the problem of optimal pricing and routing schemes for a charging network operator where users specify their priority level while the charging network operator chooses between a profit-maximizing and a social welfare-maximizing mode. The paper~\cite{le2016optimal} studies the problem of optimally charging EVs by distributing the optimization program to ultimately compute the congestion impact of a population of vehicles on the power network. In this paper, we focus on the EV charging facility pricing problem, i.e., how the choice of different pricing models affects the risk of exceeding resource capacity.

In practice, the dual purpose of EV charging facilities as both a parking location and a place to charge your vehicle will often compete against one another. From a user perspective, it is essential to have access to available charging stations, especially since many EVs have limited range. From a charging facility operator perspective, it is imperative to provide satisfactory service to users. Apart from accommodating the users, the operators may have to ensure that their total power consumption stays below a certain level to avoid overloading the power grid. Some operators have to properly manage the dual function of charging facilities serving as parking locations, with mindful considerations for space constraints and price sensitivities of users.

In this paper, we 
model the user-charging facility dynamics in a queuing framework where a user arrives with parameters which are random variables. These variables are the users' energy demand, impatience factor which quantifies how a user values their time, and their desired time at a location. The charging facility is able to charge vehicles for a fee at variable rates, and we consider two possible modes of operation: a service level model in which a user chooses from a discrete set of charging rates, and a deadline model in which a user chooses a charging deadline. In either model, the fee, i.e., price, set by the charging facility depends on the user's choice, and a user chooses a particular service level or  charging deadline which minimizes a total cost that is a combination of the actual cost to charge the vehicle and park at the facility and the user's opportunity cost from the time it takes to receive the charge. This cost is therefore a function of the user's random parameters and the prices set by the charging facility. 

The contributions of this paper are the following. First, we formalize the charging facility model and the two operating models described above. To the best of our knowledge, this is the first model of EV charging facilities that explicitly includes variable pricing for varying levels of service, parking costs, and user opportunity costs. This model provides a tractable approach for studying EV charging under diverse constraints and user assumptions. Furthermore, using
knowledge on the probability distribution of the user parameters, we derive confidence intervals on a charging facility's likelihood of not exceeding a specified number of users (i.e., occupied charging spots) and/or a specified threshold of electric power draw. Lastly, we study the practicality of the two operating models considered by demonstrating how a charging facility operator utilizes our results to set the pricing function parameters for both operating models.

This paper extends the prior works \cite{pandit2018discount,santoyo2020multi}. In particular, we expand  \cite{pandit2018discount}, which only considers the deadline model in which a user chooses a charging deadline, by considering a more general class of pricing functions, and including a cost to the user if a user desires 
to stay at a charging facility beyond their time to fully charge their vehicle. We expand \cite{santoyo2020multi}, which only considers the service level model in which a user chooses from a discrete set of charging rates, by including a parking fee.

The remainder of this paper is organized as follows: Section \ref{section: problemformulation} formally introduces relevant parameters and formulates the problem statement for the discrete and continuous pricing models, Section \ref{section: mainresults} presents the main results of this paper, i.e., probabilistic bounds on the number of users and the total power consumption for the different pricing models. Section \ref{section: casestudy} presents a case study which compares the theoretical to simulated results along with an example on how a charging facility operator can adjust the pricing parameters to stay within resource limits, and Section \ref{section: conclusion} concludes the paper. The Appendix contains proofs of some of the results presented in the paper along with some additional propositions used in the paper.

\subsection{Notation}
We denote the positive part of a real number $x$ by $[x]_+ = \max(0,x)$. For an indexed set of variables $\{x^k\}$, we let $\Delta^i_j x$ denote the difference between the variable with index $i$ and $j$, i.e., $\vdiff{i}{j}{x} = x^i - x^j$.

When considering a collection of independent and identically distributed (i.i.d) random variables indexed by subscripts, we use non-subscript variables when referring to properties that hold for any of the i.i.d random variables. For example, $\mathbb{E}[x]$ is the expectation of each i.i.d random variable $x_j$.

\section{EV Charging Facility Model Formulation}
\label{section: problemformulation}
Consider an EV charging facility that serves a local attraction or public facility, e.g., highway rest area, a shopping center, business park, hotel, or government building, and which has a finite number of parking spots which serve as parking and charging locations for individual vehicles. The charging facility therefore provides both electric energy to vehicles as well as parking.

We study the problem of how the EV charging facility should set prices for charging and parking to balance its supply of these resources with the demand of users.
 In particular, we consider two fundamental operating models:  
 in the first model, called the \textit{defined service level model} (DSL), users directly choose from a discrete set of charging rates. In the second model, called the \textit{prescribed deadline model} (PD), users indirectly choose a charging rate by specifying a departure time. In the DSL model, we assume the charging facility is able to provide electric energy at several  discrete  rates of charge for differing prices. This flexibility 
 allows the charging facility to manage both the total power usage and, indirectly, the charging facility usage. In the PD model, rather than directly choose a charging rate, 
 users 
 choose a deadline and the charging facility is assumed to provide energy at a constant rate so that the vehicle is fully charged by the deadline. 
 In both models, a user's choice is determined by the amount of charge required for their EV, the preferred amount of time they will spend at the local attraction, the prices set by the charging facility, and their impatience factor. We now make this setup and the accompanying assumptions precise.

At this facility,  a user $j$ arrives at some time $a_j$ (in hr.) with charging demand $x_j$ (in kWh), an impatience factor~$\alpha_j$ (in \$/hr.), and a desired (i.e., minimum) amount of time they will spend at the charging location $\xi_j$ (hr.). The impatience factor quantifies how much a user values their time versus money, i.e., it is 
the opportunity cost for the user to wait to charge their vehicle. Throughout the paper we will make the following assumption about the aforementioned variables.

\begin{assump}[Users]
\label{assump: rv_arrival}
User arrivals at the charging facility are a Poisson process with parameter $\lambda$ (in EVs/hr.).
Individual charging demand $x_j$, the impatience factor $\alpha_j$, and the time spent at the charging location $\xi_j$ for each user $j$ are random variables which are independent and identically distributed (i.i.d).
Additionally, 
there exists finite $0 < x_\text{min}< x_\text{max}$, $ 0\leq \alpha_\text{min}< \alpha_\text{max}$, and $0\leq \xi_\text{min}< \xi_\text{max}$ such that
the distributions of $x_j$, $\alpha_j$, and $\xi_j$ are only supported on $[x_\text{min}, x_\text{max}]$, $[\alpha_\text{min}, \alpha_\text{max}]$, and $[\xi_\text{min}, \xi_\text{max}]$, respectively.
Furthermore, each $x_j$ and $\alpha_j$ are assumed to be continuous random variables but we allow for the possibility that $\xi_{\min}=0$ and $\P(\xi_j=0)>0$ to accommodate the practical special case in which, with nonzero probability, users have no desire to remain at the charging facility. In this case, the distribution of $\xi_j$ is understood to be a generalized probability density function.
\end{assump}

\begin{table}
        \centering
        \caption{User Parameter Definitions}
        \label{table:userparam}
        \begin{tabular}{cllc} 
        \textbf{Var.} & \textbf{Parameter} & \textbf{Unit} & \multicolumn{1}{l}{\textbf{Range}} \\ \hline 
        $j$ & user index & - & - \\
        $a_j$ & arrival time & hr. & - \\
        $x_j$ & user demand & kWh & $[x_\text{min}, x_\text{max}]$\\ 
        $\alpha_j$ & impatience factor & \$/hr. & $[\alpha_\text{min}, \alpha_\text{max}]$\\
        $\xi_j$ & desired time at location & \$/hr. & $[\xi_\text{min}, \xi_\text{max}]$\\
        $r_j$ & charging rate & kW & $(0, R^\text{max}]$\\
        $u_j$ & prescribed deadline & hr  & $[u_\text{min}, u_\text{max}]$\\
        [0.1em] \hline
        \end{tabular}
        \label{table: facilityparam}
\end{table}

The user parameters, their respective units, and upper and lower bounds are summarized in Table~\ref{table:userparam}. As mentioned previously, we consider two models for how a user pays for charging their vehicle. In both models, users balance the need for electric charge with the need for a parking spot for at least their desired time at the local attraction, which may be zero. In the DSL model, a user chooses from a discrete number, possibly just one, of possible charging levels. Thus, a user can charge their vehicle faster by paying more for a higher rate of charge. In the DSL model, a user might also pay a parking fee if their vehicle reaches full charge before the user's desired time at the attraction, $\xi_j$. In the PD model, the user directly provides a departure time, i.e., a charging deadline, and the charging facility provides electric power during the resulting time window so that the vehicle has full charge at departure. In both models, a user will always remain at the charging facility at least for the desired time $\xi_j$, but they may stay longer if the charging facility offers sufficient discount for providing a slower charging rate. We formalize these two models in the next two subsections.

\subsection{Defined Service Level (DSL) Model}
In the DSL model, the charging facility offers $L$ service levels. Each service level $\ell \in \{1, \ldots, L \}$ corresponds to a distinct charging rate $R^\ell>0$ (in kW) and price $V^\ell > 0$ (in \$/kW) that is the cost per unit energy for the service level. Thus, user $j$ with energy demand $x_j$ pays $x_jV^\ell $ (in \$) to receive a full charge over the time horizon $x_j/R^\ell$ (in hr.) when choosing service level $\ell$. Additionally, the users face a parking fee $F$ (in \$/hr.) which is equal across all service levels. The parameters related to the charging facility under a discrete pricing model are listed in Table~\ref{table:facilityparam}. To distinguish the parameters related to the charging facility from those related to the users, the charging facility parameters are upper case and indexed by a superscript, while the parameters for the users are lower case and indexed by a subscript.

\begin{table}
    \centering
    \caption{Parameter Definitions for the Charging Facility under the DSL Model}
     \label{table:facilityparam}
    \begin{tabular}{cllc}
    \textbf{Var.} & \textbf{Parameter} & \textbf{Unit} & \multicolumn{1}{l}{\textbf{Range}} \\ \hline 
    $\ell$ & service level & - & $\{1,\ldots, L\}$ \\
    $V^\ell$ & price per unit of energy & \$/kWh & -\\
    $R^\ell$ & charging rate & kW & $(0, R^\text{max}]$\\
    $F$ & parking fee & \$/hr. & -\\
    [0.1em]  \hline
    \end{tabular}
\end{table}

\begin{assump}[DSL Model Charging Rates]
\label{assump: ordering_of_func}
Among $L$ service levels offered by the charging facility, a higher charging rate is more costly, i.e., if $R^i > R^k$ then $V^i > V^k$. Moreover, charging rates and prices are distinct so that $R^i \neq R^k$ for all $i \neq k$. Lastly, and without loss of generality, the charging facility's pricing functions are enumerated such that
 $V^1<V^2<\ldots <V^L$ and therefore $R^1<R^2<\ldots<R^L$. Define
the maximum charging rate $R^{\text{max}} \defineas R^L$.
\end{assump}

A user $j$ will remain at a charging facility for the amount of time to completely fulfill their demand $x_j$ and for their desired time at the local attraction, $\xi_j$, whichever is larger.  Since user $j$ values their time in excess of the time they want to spend at the charging facility at a rate $\alpha_j$, they may be willing to pay for a higher service level since it delivers a full charge faster. On the contrary, a charging facility operates under space constraints so a charging facility operator will impose a parking fee at a rate $F$ which penalizes a user for the time they spend in excess of receiving a full charge. To this end, the  total cost faced by a user arriving at the charging facility with impatience factor $\alpha_j$,  desired time spent at the charging location $\xi_j$, and charging demand $x_j$, and who chooses service level $\ell$, is given by
\begin{multline}
    \label{eq: cost_func}
        g_\ell(x_j, \alpha_j, \xi_j) = \\ x_jV^\ell  + \alpha_j\left[ \frac{x_j}{R^\ell} - \xi_j\right]_+ 
        + F \left[\xi_j - \frac{x_j}{R^\ell} \right]_+ \,.
\end{multline}
In \eqref{eq: cost_func}, the first term of the sum, $x_jV^\ell$, is the energy cost to the user resulting from their demand at arrival. The second term of \eqref{eq: cost_func}, $\alpha_j\left[ \frac{x_j}{R^\ell} - \xi_j\right]_+$, where $\frac{x_j}{R^\ell}$ is the time to charge for a particular service level~$\ell$, is the cost associated with how much a user values their time in excess of the time they sought to spend at the charging facility location. Lastly, the third term in \eqref{eq: cost_func}, $F \left[\xi_j - \frac{x_j}{R^\ell} \right]_+$, is the parking cost associated with spending more time at a charging facility than the time 
to fulfill the user's demand $x_j$. Individual users choose a service level at a charging facility which minimizes their total cost of charging factoring in their impatience. To that end, let $S(x_j, \alpha_j, \xi_j): [x_\text{min}, x_\text{max}] \times [\alpha_\text{min}, \alpha_\text{max}] \times [\xi_{min}, \xi_{max}]\to \{1, \ldots, L\}$ be defined by 
\begin{align}
    \label{eq: level_func}
    S(x_j, \alpha_j, \xi_j) =  \argmin\limits_{\ell \in \{1, \dots, L \}} g_\ell(x_j, \alpha_j, \xi_j)\,.
\end{align} 
Then, a rational user $j$ chooses service level $S(x_j, \alpha_j, \xi_j)$ in order to minimize their total cost as formalized in the later stated assumption. 

For notational convenience, we also define the values $r_j$ to be the charging rate and cost per unit of energy chosen by user $j$ after solving~\eqref{eq: level_func}, i.e., $r_j = R^{S(x_j, \alpha_j, \xi_j)}$, as indicated in Table~\ref{table:userparam}. Observe that the user charging times $x_j/r_j$, being uniquely determined by $x_j$, $\alpha_j$, and $\xi_j$, constitute a collection of independent and identically distributed random variables. Furthermore, this means the time a user spends at the charging location is $\max\left\{\xi_j, x_j/r_j\right\} $ where $x_j/r_j$ is the time for a user to receive a full charge based on their chosen service level.

\begin{assump}[DSL Users are Rational]
\label{assump: timeatcharger}
Each user chooses a charging rate according to~\eqref{eq: level_func} and leaves the charging facility once they have satisfied their charging demand or when their desired time at a charging facility has been reached, whichever is greater. Thus, user $j$ occupies a charger at the facility during the time interval $[a_j, a_j+ \max\left\{\xi_j, x_j/r_j \right\}]$. 
\end{assump}  

A practically important special case of \eqref{eq: cost_func} occurs when there is no local attraction beyond the charging facility so that users only wish to charge their vehicle, i.e., $\xi_j=0$ for all $j$ and the charging facility does not serve a secondary purpose of providing parking, and thus we may take 
$F = 0$. 
In this special case,
\eqref{eq: cost_func} becomes
\begin{align}
    \label{eq: cost_func_special_case}
    g_\ell(x_j, \alpha_j) =  x_jV^\ell  + \alpha_j\frac{x_j}{R^\ell} \, .
\end{align}
We refer to this
special case as 
the \textit{DSL free parking model} (DSL-FP model). If we refer to the DSL model in an instance which excludes the DSL-FP special case, we sometimes refer to it 
as the \textit{DSL metered parking model} (DSL-MP model) for emphasis. 

Note that the DSL-FP model is the focus of our prior work~\cite{santoyo2020multi};
hence, the cost function \eqref{eq: cost_func} is a generalization of that considered in~\cite{santoyo2020multi} which accounts for parking fees and users staying at a particular location longer than their time-to-charge. In the DSL-FP model, as in the DSL-MP model, 
given a collection of pricing functions as in \eqref{eq: cost_func_special_case}, a user $j$ chooses their service level by solving~\eqref{eq: level_func}, but we omit $\xi_j$ as an argument in the corresponding expressions. 

As described above, the DSL model allows for a charging facility that offers multiple discrete charging levels. For example, this model is well-suited for existing charging infrastructure, which is currently divided into three charging levels \cite{energygov}.  

However, it may be more convenient for the user to provide a deadline by which they expect to receive a full charge, and for the charging facility to determine a price and charge rate to fulfill this deadline. Such pricing schemes have indeed been implemented in practice~\cite{nrel}. In other words, it may be the case that the users are not restricted to a predefined set of service levels, but rather can pick any charging rate by proxy of choosing a charging deadline. This setting is characterized in the following subsection.

\subsection{Prescribed Deadline (PD) Model}
In this section, we construct the PD model for the EV charging facility and utilize the variable definitions presented in Table \ref{table: facilityparam}. 

As in the DSL model, in the PD model 
a user $j$ arrives with charging demand $x_j$, a desired time at the location $\xi_j$, and an impatience factor $\alpha_j$. However, in the PD model, the user $j$ chooses a charging deadline $u_j$ rather than a discrete charging rate.  The charging facility broadcasts a single pricing function $P(x_j, u_j)$ 
that constitutes the financial cost to a user receiving charge~$x_j$ over the deadline $u_j$.  
Then, 
\begin{align}
    \label{eq: total_cost_func_continuous}
    C(x_j, u_j, \alpha_j, \xi_j) = P(x_j, u_j) + \alpha_j (u_j - \xi_j) \, 
\end{align}
is the total cost faced by a user $j$ who arrives with demand $x_j$, impatience factor $\alpha_j$, planned time at location $\xi_j$, and who chooses a charging deadline $u_j\geq \xi_j$. Hence,~\eqref{eq: total_cost_func_continuous}, penalizes users more for choosing a charging deadline $u_j$ greater than their desired time at a location~$\xi_j$ at a rate $\alpha_j$. 
A rational user $j$ chooses their  charging deadline according to
\begin{align} \label{eq: continuousargmin}
    u_j \in \argmin_{u \geq \xi_j} C(x_j, u , \alpha_j, \xi_j) \, .
\end{align}
In \eqref{eq: total_cost_func_continuous}, we see that in addition to paying a price to charge as a function of the demand and chosen deadline, a user faces an opportunity cost which is a function of their impatience and how much time beyond their desired time,~$\xi_j$, they spend at the charging facility.
\begin{assump}[PD Users are Rational]
\label{assump: timeatcharger_continuous}
Each user chooses a charging deadline according to~\eqref{eq: continuousargmin} and leaves the charging facility at the chosen deadline. 
Thus, user $j$ occupies a charger at the facility during the time interval $[a_j, a_j+ u_j]$.
\end{assump} 

We explore the problem of charging facilities within the context of limited resources; therefore, there are physical limitations on the charging facilities such as a maximum charging rate allowable per user. This point is formalized in the following assumption. 

\begin{assump}[PD Model Charging Rates]
\label{assump: chargeratevaluerange}
The pricing function $P(x_j,u_j)$ is such that there exists  an upper bound $R^\text{max}$ on the charging rate for any user solving \eqref{eq: continuousargmin} under the PD model, \emph{i.e.}, 
$R^\text{max} \geq r_j \,$, where $r_j = x_j/u_j$, for all users $j$ when $u_j$ is chosen according to \eqref{eq: continuousargmin}. Moreover, 
the charging facility provides  electric power at the constant rate $r_j$ over the charging time horizon $u_j$ for each user $j$. 
\end{assump}
\begin{remark}
In the PD model, note that the 
charging deadline $u_j$, and therefore also the
charging rate $r_j=x_j/u_j$, 
is a continuous random variable. This contrasts with the DSL model where $r_j$ is a discrete random variable.
\end{remark}

There exist many candidate functions that can be utilized as pricing functions. Analysis of the minimizer in~\eqref{eq: continuousargmin} is particularly amenable in the case that the pricing function $P(x_j, u_j)$ is convex in the deadline variable~$u_j$. 
In that case, $C(x_j, u_j, \xi_j)$ is also convex in $u_j$, and hence there exists a unique minimizer $u^*$ for \eqref{eq: continuousargmin}. 
The following is an example of such a pricing function.
\begin{example}
\label{ex:1}
Consider the pricing function
\begin{align}
    \label{eq: pricingfunc_continuous}
    P(x_j, u_j) = x_j \left(D(u_j - \omega)^2 + B\right) \,,
\end{align}
where $D$ is the surge price (in \$/kWh-hr.\textsuperscript{2}), 
$\omega$ is the charging facility's desired time spent at the charger (in hr.), 
and $B$ is the base price (in \$/kWh).
Then, from \eqref{eq: continuousargmin}, a user $j$ chooses deadline
\begin{align}
    \label{eq: continuousargmin_example}
    u_j \in \argmin_{u \geq \xi_j} \, \,  x_j \left(D(u - \omega)^2 + B\right) + \alpha_j(u - \xi_j)\, .
\end{align}
Note that the term $\left(u_j - \omega \right)^2$ allows the charging facility to penalize a user for choosing a deadline less than or greater than $\omega$  at a surge price rate $D$. The surge price penalty is in addition to the base price $B$ that a user pays for their charging demand.

Since \eqref{eq: total_cost_func_continuous} substituted with \eqref{eq: pricingfunc_continuous} is convex in $u$, 
the minimizer is unique and available in closed-form so that user $j$ will choose deadline
\begin{align}
    \label{eq: minimizer_example}
    u_j \defineas u^* = \max\left\{\xi_j, \frac{-\alpha_j }{2 D x_j} + \omega \right\}.
\end{align}
As previously mentioned, we operate under Assumption~\ref{assump: chargeratevaluerange}, i.e., $R^\text{max} \geq x_j/u_j$ must hold. Interpreting $R^\text{max}$ as an \emph{a priori} fixed limit, the charging facility must then choose parameters $D$, $B$, and $\omega$ to satisfy Assumption \ref{assump: chargeratevaluerange}. Algebraic manipulations combined with reasoning when the maximum is attained lead to the fact that $\omega  > x_\text{max}/R^{max}$ and 
\begin{equation}
D  >   \left[ \max_{x_j\in [x_{\text{min}}, x_{\text{max}}]} \frac{\alpha_{\text{min}} R^{max}}{2\omega  x_{j}R^{max} -  2 x_{j}^2}\right]_+ \,. \label{eq:D}
\end{equation}

In practice, \eqref{eq:D} provides a 
way for charging facilities to set the surge price $D$ so that the charging rate limit $R^\text{max}$ for each user is satisfied.
\end{example}


    
    \begin{remark}
    \label{remark: chargingtimedistinction}
     In the DSL model the user will remain at the charging location for $ \max\{\xi_j, x_j/r_j \}$. This means that there exists the possibility that the user selects a charging rate which fulfills the vehicles charging demand before the user has reached their desired time to spend at the location~$\xi_j$. However, in the PD case, 
    a user selects a deadline $u_j$ and the appropriate rate is set 
    that fulfills the charging demand exactly at the deadline time.
    This means that in the PD model the vehicle will finalize charging exactly when the users leaves.
    \end{remark}  
    
     Remark \ref{remark: chargingtimedistinction} points to a subtle distinction that is important because we are interested in providing probabilistic bounds on the number of present users at the charging facility. Hence, in the DSL model, there is a difference in the number of users actively charging and the number of users present at the charging facility.
    
    Next, we formally introduce the problem statement for the charging facility which lays the foundation for the main result for both the DSL and PD models and which captures the subtle distinction in Remark \ref{remark: chargingtimedistinction}.

\subsection{Guarantees on Charging Facility Capacity Limits}
Charging facilities are concerned with adhering to both user capacity and energy consumption restrictions. Let the set of present users at the charging facility at time $t$ be defined as 
\begin{equation*}
N(t) = 
    \begin{cases}
        \{ i : t \in [a_{i}, a_{i} + \max\{\xi_i, \frac{x_i}{r_i}\}] \}  \quad &\text{if DSL} \\
        \{ i : t \in [a_{i}, a_{i} + u_{i}] \} \quad &\text{if PD} 
    \end{cases}
\end{equation*}
 and let $\eta(t)=|N(t)|$ be the cardinality of the set of present users. Moreover, let the set of actively charging users be
 \begin{equation*}
N_{\text{act}}(t) = 
    \begin{cases}
        \{ i : t \in [a_{i}, a_{i} + \frac{x_i}{r_i}] \}  \quad &\text{if DSL} \\
        \{ i : t \in [a_{i}, a_{i} + u_{i}] \} \quad &\text{if PD} 
    \end{cases}
\end{equation*}
and let $\eta_{\text{act}}(t)=|N_\text{act}(t)|$ be the cardinality of the set of actively charging users. Then, 
 \begin{equation*}
Q(t) = \sum\limits_{i \in N_{\text{act}}(t)} r_{i} 
\end{equation*}
is the total charging rate at time $t$ for all actively charging users, i.e., the charging facility's total power consumption. Note that $r_i = x_i/u_i$ in the summation for the PD model. \par 

We consider the problem in which the charging facility is interested in providing probabilistic guarantees on the number of present users in the system and the total power requirements at any given time $t$. We thus  wish to compute a high-confidence bound on the total number of active users and their respective aggregate power draw at any given time, as is made precise in the following problem statement. 
\begin{probstatement}
\label{probstat: problem_statement}
Given an EV charging facility operating under the DSL (resp., PD) model satisfying Assumption \ref{assump: ordering_of_func}  and \ref{assump: timeatcharger} (resp., Assumption \ref{assump: timeatcharger_continuous} and \ref{assump: chargeratevaluerange}) and EV users satisfying Assumption \ref{assump: rv_arrival}, for any $\mathcal{M}$ number of users and $\mathcal{R}$ total charging facility power consumption rate, compute $\delta(\mathcal{M})$ and $\gamma(\mathcal{R})$ such that
\begin{equation}
    \label{eq: prob_M}
    \P(\eta(t)< \mathcal{M} ) \geq 1 - \delta(\mathcal{M})
\end{equation}
and
\begin{equation}
    \label{eq: prob_R}
    \mathbb{P}(Q(t)< \mathcal{R}) \geq 1 - \gamma(\mathcal{R}) \, .
\end{equation}
\end{probstatement}

\section{Main Results}
\label{section: mainresults}
In this section, we first introduce several propositions which formalize the probability that a randomly selected user will choose a particular service level in the DSL model.
 These results elucidate the remarkable fact that, conditioned on the ratio $x_j/\xi_j$, the probability of choosing a particular rate in the DSL model depends only on the impatience factor $\alpha_j$. 
We formalize the results for the DSL model in the following subsection. We do not present similar results for the PD model since many of the distributions of interest in the PD model are derived distributions resulting from algebraic operations on random variables and thus, in general, do not have closed form expressions. Lastly, we present a theorem which solves the problem statement above and provides probabilistic guarantees of the form \eqref{eq: prob_M}--\eqref{eq: prob_R} for both the DSL and PD model. 
\par 

\subsection{User Choice under DSL Model }
\label{subsec: discretepricing_probabilityoflevel}
Define
the ratio
\begin{align*}
    \rho_j = x_j/\xi_j \,.
\end{align*}
In the case that $\xi_j=0$, which by Assumption \ref{assump: rv_arrival} may occur with nonzero probability, we take $\rho_j=\infty$ and the following analysis still holds.
When analyzing the pricing function in \eqref{eq: cost_func}, it becomes apparent that a user $j$ selecting service level $\ell$ pays either a cost associated with their opportunity cost (i.e., impatience) when $x_j/R^\ell > \xi_j$ (equivalently, $\rho_j>R^\ell$) or  a cost associated with the parking fee when $x_j/R^\ell < \xi_j$ (equivalently,  $\rho_j<R^\ell$). 
Note that $\rho_j$ (in kW) is the charging rate that would deliver full charge to user $j$ over their desired time at location $\xi_j$.

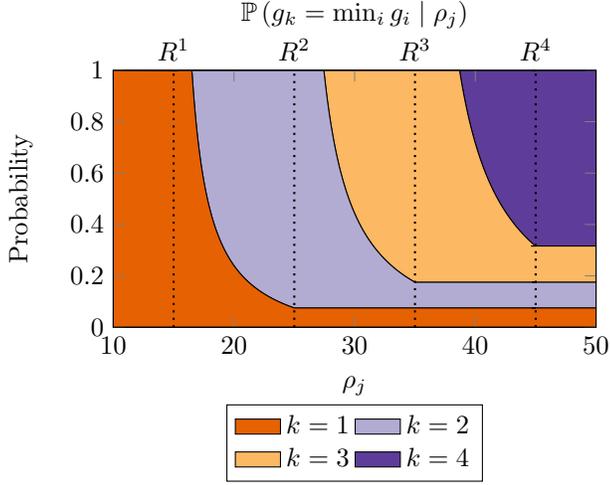
\begin{figure}
    \centering
    \input{plots/fig1.tikz}
    \caption{
    The figure shows the probability that a user $j$ with desired charging rate $\rho_j = x_j/\xi_j$ chooses service level $k$. Depending upon the users impatience factor, a user may choose a lower charging level than the partition boundaries of where   the desired charging rate~$\rho_j$ lies.
    }
    \label{fig: rho_bins}
\end{figure}

 The random variable $\rho_j$ has domain $[\rho_{\text{min}}, \rho_{\text{max}}]$ with $\rho_{\text{min}}=x_{\text{min}}/\xi_\text{max}$ and $\rho_{\text{max}}=x_{\text{max}}/\xi_\text{min}$ with possibly $\rho_{\text{max}}=\infty$. Moreover, since $\rho_j$ is user $j$'s desired charge rate, we partition its domain based on the $L$ service levels. In particular, we obtain the partition intervals $\rho_j < R^1$, $\rho_j\in \left[ R^{m} , R^{m+1} \right)$ for all $m \in \{1, \dots ,L-1\}$, and $R^L \leq \rho_j$. This is illustrated in Fig.~\ref{fig: rho_bins}. As a result of this observation, we can find the probability that a particular service level will be the minimizing service level  while conditioning on  $\rho_j$ and evaluating the conditional probability at each of the aforementioned intervals. 

We formalize these observations in Proposition \ref{proposition: probability_choose_level} which defines the probability the DSL pricing functions of the form of \eqref{eq: cost_func} will be the minimum within the set of service levels and hence will be the choice of a particular user. For the remainder of the paper we let 
$\bar{R}_\ell = 1/R_\ell$ for all $\ell\in\{1,\ldots,L\}$. Additionally,
$f_\Rho(\rho)$ is the distribution of~$\rho_j$ supported on  $[\rho_\text{min}, \rho_\text{max}]$ and note that $\E_\Rho\left[ z(\rho) \right] = \int^{\infty}_{-\infty} z(\phi)f_\Rho(\phi) \d\phi$ for some function $z(\rho)$.


\begin{prop}
    \label{proposition: probability_choose_level}
     Under Assumptions \ref{assump: rv_arrival}, \ref{assump: ordering_of_func}, and \ref{assump: timeatcharger}, consider the set of $L$ functions of three independent RVs $\big\{g_\ell(x_j,\alpha_j, \xi_j) \big\}^L_{\ell = 1}$ where each $g_\ell$ is of the DSL model as defined in \eqref{eq: cost_func} and selection function $S(x_j,\alpha_j,\xi_j)$ as defined in \eqref{eq: level_func}. Then,  for $k \in \{1, \ldots, L \}$,
    \begin{multline*}
         \P(S(x_j, \alpha_j ,\xi_j) = k ) = \E_\Rho\left[\P\left( g_k = \min_{i} g_i\mid \rho_j \right)\right]\,,
    \end{multline*}
     
    where 
    \begin{multline*}
    \label{eq: condminprobabilitycases}
    \P\left( g_k = \min_{i} g_i\mid \rho_j \right) = \\
    \begin{cases}
    1 & \text{if }\rho_j < R^1 \land k = 1\\
    0 & \text{if }\rho_j < R^1 \land k > 1\\ 
    \P\left(\ubar{\alpha}^k_1 < \alpha_j < \Bar{\alpha}^k_1 \right) & \text{if } \rho_j \in [R^m , R^{m+1}) \land k \leq m \\[0.5em]
    \P\left(\ubar{\alpha}^k_2 < \alpha_j < \alpha_\text{max} \right) & \text{if } \rho_j \in [R^m , R^{m+1})  \land   k= m+1 \\
    0 &\text{if } \rho_j \in [R^m , R^{m+1}) \land  k > m+1 \\
    \P\left(\ubar{\alpha}^k_3 < \alpha_j < \Bar{\alpha}^k_3 \right) &\text{if } \rho_j \geq R^L 
    \end{cases}
    \end{multline*}
    and
    \begin{align*}
        \Bar{\alpha}^k_1 &=\\  
        &\hspace{-0.4cm}\min\left(\alpha_\text{max}, \min_{m < i} \frac{F \left(\frac{1}{\rho_j} - \frac{1}{R^{i}}\right) - \vdiff{k}{i}{V}}{\frac{1}{R^k} - \frac{1}{\rho_j}}, \min _{k < i \leq m}\frac{\vdiff{i}{k}{V}}{\vdiff{k}{i}{\bar R}}\right)\,, \\
        \ubar{\alpha}^k_1 &= \max\left(\alpha_\text{min}, \max_{i < k \leq m}\frac{\vdiff{i}{k}{V}}{\vdiff{k}{i}{\bar R}}\right)\,, \\
        \ubar{\alpha}^k_2 &= \max\left(\alpha_\text{min}, \max_{i < m + 1}\frac{F \left(\frac{1}{R^k} - \frac{1}{\rho_j} \right) - \vdiff{k}{i}{V} }{\frac{1}{\rho_j} - \frac{1}{R^i}}\right) \,, 
        \\
        \Bar{\alpha}^k_3 &= \min\left(\alpha_\text{max}, \ \min_{k < i}\frac{\vdiff{i}{k}{V}}{\vdiff{k}{i}{\bar R}}\right)\,,
        \\
        \ubar{\alpha}^k_3 &= \max\left(\alpha_\text{min},\ \max_{i<k} \frac{\vdiff{i}{k}{V}}{\vdiff{k}{i}{\bar R}}\right)\,.
    \end{align*}
    
     Furthermore, the charging rates $r_j$ chosen by each user $j$ is a collection of independent and identically distributed discrete random variables each with probability mass function (PMF)
    \begin{align}
        p_{r}(r) = 
        \begin{cases}       
            \E_\Rho\left[\P\left( g_1 = \min_{i} g_i\mid \rho_j \right)\right] & \text{if }  r = R^1 \,, \\ 
            \qquad \qquad \qquad  \vdots \quad \quad  \\
            \E_\Rho\left[\P\left( g_L = \min_{i} g_i\mid \rho_j \right)\right] & \text{if } r = R^L \,.
        \end{cases}
    \end{align}
\end{prop}

Proposition \ref{proposition: probability_choose_level} states that the probabilities of a service level being the minimizing choice for a user is computable by leveraging the law of total probability. 
While in some intervals, the  probability $\P\left( g_k = \min_{i} g_i\mid \rho_j \right)$ is either $0$ or $1$, in others it is an integration over an interval of the distribution of the impatience factor where there exists a possibility that $\P\left( g_k = \min_{i} g_i\mid \rho_j \right)$ is zero within a particular partition. Lastly, Proposition \ref{proposition: probability_choose_level} formalizes the fact that choosing a service level amounts to choosing a particular charging rate and presents the resulting PMF for the charging rates $r_j$.  
The  proof of Proposition \ref{proposition: probability_choose_level}
is in Appendix \ref{app: propmin}.

Fig. \ref{fig: rho_bins} illustrates $\P\left( g_k = \min_{i} g_i\mid \rho_j \right)$ when $L=4$ service levels. Considering the case when $\rho_j < R^1$, we see that $\P\left( g_1 = \min_{i} g_i\mid \rho_j \right) = 1$ which follows the first case statement for $\P\left( g_k = \min_{i} g_i\mid \rho_j \right)$ in Proposition~\ref{proposition: probability_choose_level}. Analyzing the case when $\rho_j \in [R^1, R^2)$, we see that $\P\left( g_1 = \min_{i} g_i\mid \rho_j \right)$ decreases while $\P\left( g_2 = \min_{i} g_i\mid \rho_j \right)$ increases for increasing $\rho_j$. Lastly, considering the case when $\rho_j \in [R^2, R^3)$, we see that $\P\left( g_1 = \min_{i} g_i\mid \rho_j \right)$ is constant, $\P\left( g_2 = \min_{i} g_i\mid \rho_j \right)$ decreases, and $\P\left( g_3 = \min_{i} g_i\mid \rho_j \right)$ increases for increasing $\rho_j$. Note however that while $\rho_j \in [R^2, R^3)$, it may still be optimal for a user to choose level $R^1$. This results from the impatience factor with which a user arrives. Specifically, if a users impatience factor is low enough they may opt to choose a slower charging rate since the parking fee may result in a higher cost. Note that the same phenomenon occurs when $\rho_j \in [R^3, R^4)$ and extends for any $L$ service levels.


In the DSL-FP special case described by the total cost function \eqref{eq: cost_func_special_case}, Proposition \ref{proposition: probability_choose_level} becomes the following corollary.
\begin{cor}
    \label{corollary: probability_choose_level_special_case}
     Under Assumptions \ref{assump: rv_arrival}, \ref{assump: ordering_of_func}, and \ref{assump: timeatcharger}, consider the set of $L$ functions of two independent RVs $\big\{g_\ell(x_j,\alpha_j) \big\}^L_{\ell = 1}$ where each $g_\ell$ is of the DSL-FP model  as defined in \eqref{eq: cost_func_special_case}. Then,  for $k \in \{1, \ldots, L \}$,
    \begin{equation*}
         \P\bigl(S(x_j, \alpha_j) = k \bigr) = \P \left(\ubar{\alpha}^k < \alpha_j < \Bar{\alpha}^k \right)
    \end{equation*}
    where 
    \begin{align}
        \label{eq:alpha_1}
        \Bar{\alpha}^k &= \min\left(\alpha_\text{max}, \ \min_{k < i}\frac{\vdiff{i}{k}{V}}{\vdiff{k}{i}{\bar R}}\right),\\
        \label{eq:alpha_2}\ubar{\alpha}^k &= \max\left(\alpha_\text{min},\ \max_{i<k} \frac{\vdiff{i}{k}{V}}{\vdiff{k}{i}{\bar R}}\right).
    \end{align}
    Furthermore, the charging rates $r_j$ chosen by each user $j$ is a collection of independent and identically distributed discrete random variables each with PMF 
    \begin{equation}
        \label{eq: pmf_special_case}
        p_{r}(r) = 
        \begin{cases}       
            \P \left(\ubar{\alpha}^1 < \alpha_j < \Bar{\alpha}^1 \right) & \text{if } r = R^1 \,, \\ 
            \qquad \quad  \vdots &  \\
            \P \left(\ubar{\alpha}^L < \alpha_j < \Bar{\alpha}^L \right) & \text{if } r = R^L \,.
        \end{cases}
    \end{equation}
\end{cor}

\begin{pf}
The DSL-FP special case with total cost functions of the form of \eqref{eq: cost_func_special_case} arises with $F = 0$ and $\xi_j = 0$. Simply substituting these values into the general DSL model implies that the desired charge rate  $\rho_j = \infty$ for all users~$j$. In practice, this means that users arrive desiring to charge as fast as possible. In return, this means that Corollary \ref{corollary: probability_choose_level_special_case} is just the application of the $\rho_j > R^L$ case of $ \P\left( g_k = \min_{i} g_i\mid \rho_j \right)$ in Proposition \ref{proposition: probability_choose_level}.
\end{pf}

Corollary \ref{corollary: probability_choose_level_special_case} states that, when a given service level is chosen  with nonzero probability, there exists an interval of impatience factor values for which that service level minimizes the total cost to a user. The probability that the given service level will be chosen is therefore computed by integrating the distribution of $\alpha_j$ on that interval.
This special case is the focus of our previous work \cite{santoyo2020multi}.

Proposition \ref{proposition: probability_choose_level} and Corollary \ref{corollary: probability_choose_level_special_case} define the probability a user will choose a particular service level under the DSL model. As noted, this probability is equivalent to the probability of choosing a specific charging rate. Since choosing a charging rate is a discrete choice, an explicit PMF is  available for the rates of charge. This fact will be used for the main result in the next section.

\subsection{High-Confidence Bounds on Resource Usage}
To state the high-confidence bounds on the total number of vehicles at the charging facility along with the aggregate power consumption, we start by making some observations about the expected charging rate and expected deadline for the different models.

In the DSL model $\mathbb{E}[r]=\sum^L_{\ell = 1}R^\ell p_r(R^\ell)$ and $\mathbb{E}[r^2]=\sum^L_{\ell = 1}\left(R^\ell\right)^2 p_r(R^\ell)$. In the DSL model, computing the probability a random user chooses a particular service level is an integration over the distribution of the impatience factor $\alpha_j$; however, there exists a difference in this computation between the DSL-MP and DSL-FP models. Specifically, the choice of charging rate $r_j$ chosen by a user $j$ is only a function of the impatience factor $\alpha_j$ in the DSL-FP model. Thus in the DSL-FP model $r_j$ is independent of $x_j$ so that $\mathbb{E}[x/r]={\mathbb{E}[x]}{\mathbb{E}[1/r]}$
is the expected charging time for each user $j$. In the DSL-MP model, there arises a dependency between $r_j$ and $x_j$ as a result of the ratio $\rho_j = x_j/\xi_j$ appearing in the integration bounds of Proposition \ref{proposition: probability_choose_level}. Hence, to compute $\mathbb{E}[x/r]$ one has to find the derived distribution of the ratio $x_j/r_j$ rather than simply dividing the expectations as is the case in the DSL-FP model. 

In the PD model, the distribution of $r_j$ can be found by finding the derived distribution of the ratio $x_j/u_j$ and depends on the distributions of $x_j$ and $u_j$. The deadline $u_j$ is also a function of random variables, e.g., \eqref{eq: minimizer_example} in the case when the pricing function is \eqref{eq: pricingfunc_continuous}, and hence its distribution is also a derived distribution from the distributions of $\alpha_j$ and $x_j$. Once the distributions of $u_j$ and $r_j$ are computed, one can compute $\E[u]$, $\E[r]$, and $\E[r^2]$ for the continuous pricing scheme which are of interest in the main result of this paper.
\par 

Next, we state the main theorem for this paper which addresses Problem Statement \ref{probstat: problem_statement} for both the DSL and PD models.
\begin{thm} 
\label{theorem: mainresult}
Consider a charging facility operating under the DSL model (resp., PD model) with 
Assumptions \ref{assump: rv_arrival}, \ref{assump: ordering_of_func}, and \ref{assump: timeatcharger} (resp., Assumptions \ref{assump: rv_arrival}, \ref{assump: timeatcharger_continuous}, and \ref{assump: chargeratevaluerange}).  Let
\begin{align*}
    \theta = 
    \begin{cases}
    \max\left\{\xi, x/r\right\} \ &\text{if DSL model} \\
    u \ &\text{if PD model},
    \end{cases}
\end{align*}
i.e., the model dependent random time spent at the charging facility for each user, and let 
\begin{align*}
    \theta_{act} = 
    \begin{cases}
     x/r \ &\text{if DSL model} \\
    u \ &\text{if PD model},
    \end{cases}
\end{align*}
i.e., the model dependent random time spent actively charging at the charging facility for each user.
Given any $\mathcal{M} \geq 0$ number of users and $\mathcal{R} \geq 0$ total charging rate, the following  statements hold at steady state for any time~$t$:
    \begin{enumerate}
        \item With confidence $1 - \delta(\mathcal{M})$, where
        \begin{multline*}
           \hspace{-20pt} \delta\left(\mathcal{M}\right) =\\  
           \hspace{-20pt} \begin{cases}
            \exp{\left(\frac{-(\mathcal{M}-\lambda \E[\theta])^2}{2\left(\lambda \E[\theta] + \frac{(\mathcal{M}-\lambda \E[\theta])}{3}\right)}\right)} & \text{if } \mathcal{M} > \lambda \E[\theta]  \\
            1 & \text{otherwise,}
            \end{cases}
        \end{multline*}
        the number of users will not exceed $\mathcal{M}$, i.e., $\mathbb{P}(\eta(t) < \mathcal{M} ) \geq 1 - \delta(\mathcal{M})$.
        \item With confidence $1 - \gamma(\mathcal{R})$, where 
        \begin{multline*}
            \hspace{-20pt}\gamma\left(\mathcal{R}\right) = \\
            \hspace{-20pt}\begin{cases}
            \min\Bigg\{ 1, 
            \sum_{m=\ceil[\big]{\frac{\mathcal{R}}{R^{\text{max}}}}}^{\floor[\big]{\frac{\mathcal{R}}{\E\left[r\right]}}} \exp{\left(\frac{-\left(\mathcal{R} - m\E[r]\right)^2}{2\left(m\E[r^2] + \frac{R^\text{max}\left(\mathcal{R} - m\E[r]\right)}{3}\right)}\right)} \\
            \times\mathbb{P}\bigl(\eta(t) = m \bigr) + \delta_\text{act}\left(\floor[\bigg]{\frac{\mathcal{R}}{\E\left[r\right]}}\right)\Bigg\},  \quad \text{if }   \mathcal{R} > \lambda\E\left[\theta_\text{act}\right]\E\left[r\right] 
        \\
        1,  \qquad \text{otherwise,}
            \end{cases}
        \end{multline*}
        and 
        \begin{multline*}
           \hspace{-20pt} \delta_{act}\left(\mathcal{M}\right) =\\  
           \hspace{-20pt} \begin{cases}
            \exp{\left(\frac{-(\mathcal{M}-\lambda \E[\theta_{act}])^2}{2\left(\lambda \E[\theta_\text{act}] + \frac{(\mathcal{M}-\lambda \E[\theta_{act}])}{3}\right)}\right)}, & \text{if } \mathcal{M} > \lambda \E[\theta_{act}]  \\
            1 & \text{otherwise,}
            \end{cases}
        \end{multline*}
        the total charging rate for all active users will not exceed $\mathcal{R}$, i.e., $\mathbb{P}\left(Q(t)< \mathcal{R}\right) \geq 1 - \gamma(\mathcal{R})$.
    \end{enumerate}
\end{thm}
Note that $\delta_\text{act}\left(\mathcal{M}\right)$ is very similar to $\delta\left(\mathcal{M}\right)$ in that like $\delta\left(\mathcal{M}\right)$ it is used for providing a confidence interval on $\mathcal{M}$; however, note that $\delta_\text{act}\left(\mathcal{M}\right)$ is used for providing confidence on the number of users actively charging rather than those solely present at the charging facility. 
\begin{pf}
    \begin{enumerate}
        \item 
        We begin by proving the first statement.
        First, we make use of Proposition~\ref{prop: BernoullitoRV} in Appendix~\ref{sec:appendix_obs} stating that for a Poisson random variable $Z$ with mean $\bar\lambda > 0$, and for any $\mathcal{M} >
        \bar\lambda$, $\P \bigl({Z} < \mathcal{M} \bigr) \geq 1 - \delta^\dagger(\mathcal{M})$ where 
        \begin{equation}
                \label{eq: delta_dagger_in_proof}
                \delta^\dagger(\mathcal{M}) = \exp{\left(\frac{-\left(\mathcal{M} - \bar\lambda \right)^2}{2\left(\bar\lambda + \frac{\mathcal{M} - \bar\lambda}{3}\right)}\right)}\,.
        \end{equation} 
       Since the arrival and service process can be seen as an $M/G/\infty$ queue, $\eta(t)$ is itself a Poisson random variable for each $t$ with mean $\lambda \E[\theta]$ \cite[Equation (9)]{massey2002analysis} and hence letting $\bar \lambda = \E[\eta(t)] = \lambda \E[\theta]$ proves the first case of the statement. For the second case, observe that if $\mathcal{M} < \bar\lambda$,  Proposition~\ref{prop: BernoullitoRV} cannot be applied and hence $\delta(\mathcal{M}) = 1$ gives a trivial bound for the sought probability.

        
        \item 
        Introduce $\nu$ as $\nu = \mathcal{R} - \eta_\text{act}(t)\E[r]$. Hence $\mathcal{R} = \eta_\text{act}(t)\E[r] + \nu$. By total probability, it holds that 
        \begin{multline} \label{eq:Qsum}
        \P\bigl(Q(t) \geq  \mathcal{R} \bigr) \\
        =  \sum_{m = 0}^{\infty} \P\bigl(Q(t) \geq  \mathcal{R} \mid  \eta_\text{act}(t) = m\bigr)\mathbb{P}\bigl(\eta_\text{act}(t) = m \bigr) \,.
         \end{multline}
        Next we observe that the probability that $Q(t) \geq \mathcal R$ is zero when $m < \mathcal R/R^\text{max}$. This since even if all users choose the maximum rate, it is impossible that $Q(t)$ exceeds $\mathcal R$, i.e., $\P\bigl(Q(t) \geq  \mathcal{R} \mid  \eta_\text{act}(t) = m\bigr) = 0$ for $m < \mathcal R/R^\text{max}$. Using this fact and expanding \eqref{eq:Qsum} for some $\kappa > \mathcal R/R^\text{max}$ gives
        \begin{multline} 
        \label{eq: QsumI}
        \P\bigl(Q(t) \geq  \mathcal{R} \bigr) = \\
          \sum_{m = \ceil[\big]{\frac{\mathcal{R}}{R^{\text{max}}}}}^{\kappa} \P\bigl(Q(t) \geq  \mathcal{R} \mid  \eta_\text{act}(t) = m\bigr)\mathbb{P}\bigl(\eta_\text{act}(t) = m \bigr) \\+ \sum_{m = \kappa+1}^{\infty} \P\bigl(Q(t) \geq  \mathcal{R} \mid  \eta_\text{act}(t) = m\bigr)\mathbb{P}\bigl(\eta_\text{act}(t) = m \bigr)  \,.
         \end{multline}
        For $\kappa < m < \infty$, using the fact that $\P\bigl(Q(t) \geq  \mathcal{R} \mid  \eta_\text{act}(t) = m\bigr) \leq 1$ and that $\mathbb{P}\bigl(\eta_\text{act}(t) > \kappa \bigr) = \sum_{m = \kappa+1}^{\infty}\mathbb{P}\bigl(\eta_\text{act}(t) = m \bigr)$, \eqref{eq: QsumI} becomes 
        \begin{multline} \label{eq:QsumII}
        \P\bigl(Q(t) \geq  \mathcal{R} \bigr) \leq \\ 
         \sum_{m = \ceil[\big]{\frac{\mathcal{R}}{R^{\text{max}}}}}^{\kappa} \P\bigl(Q(t) \geq  \mathcal{R} \mid \eta_\text{act}(t) = m\bigr)\mathbb{P}\bigl(\eta_\text{act}(t) = m \bigr) \\ + \mathbb{P}\left(\eta_\text{act}(t) > \kappa \right) \,.
        \end{multline}
        For a fixed $m$, and hence fixed $\eta_{act}$, such that $\nu \geq 0$, Fact~\ref{defintion-fact: BernsteinsIneq} (Bernstein's Inequality) in Appendix~\ref{sec:appendix_obs} can be utilized with $b = R^\text{max}$ and $n = \eta_\text{act}(t)$, so that each summation term in~\eqref{eq:QsumII} is bounded as
    \begin{multline*}
        \P \bigl(Q(t) \geq  \eta_\text{act}(t)\E[r] + \nu \mid \eta_\text{act}(t)  \bigr) \\
        \leq \exp{\left(\frac{-\nu^2}{2\left(\eta_\text{act}(t)\E[r^2] + \frac{R^\text{max}\nu}{3}\right)}\right)}\,.
    \end{multline*}
    Note that to apply Bernstein's inequality, $\nu \geq 0$, which is equivalent to $\mathcal{R} - m\E\left[r\right] > 0$. This implies $m < \mathcal{R}/\E\left[r\right]$. Hence, we choose $\kappa = \floor[\big]{\mathcal{R}/\E\left[r\right]}$, i.e., the floor value of $\mathcal{R}/\E\left[r\right]$.
    Using the above facts, \eqref{eq:QsumII} can be rewritten as
    \begin{multline*}
        \P\bigl(Q(t) \geq  \mathcal{R}\bigr) \\
        \leq  \sum_{m = \ceil[\big]{\frac{\mathcal{R}}{R^{\text{max}}}}}^{ \floor[\big]{\frac{\mathcal{R}}{\E\left[r\right]}}} \exp{\left(\frac{-\left(\mathcal{R} - m\E[r]\right)^2}{2\left(m\E[r^2] + \frac{R^\text{max}\left(\mathcal{R} - m\E[r]\right)}{3}\right)}\right)} \\ \times 
        \mathbb{P}\bigl(\eta_\text{act}(t) = m \bigr)  + \mathbb{P}\left(\eta_\text{act}(t) >  \floor[\big]{\mathcal{R}/\E\left[r\right]} \right)\,.
    \end{multline*}
   Now, by utilizing the result from Statement 1 of Theorem \ref{theorem: mainresult}, we obtain
    \begin{multline*}
        \P\bigl(Q(t) \geq  \mathcal{R}\bigr) \\
        \leq \sum_{m = \ceil[\big]{\frac{\mathcal{R}}{R^{\text{max}}}}}^{\floor[\big]{\frac{\mathcal{R}}{\E\left[r\right]}}} \exp{\left(\frac{-\left(\mathcal{R} - m\E[r]\right)^2}{2\left(m\E[r^2] + \frac{R^\text{max}\left(\mathcal{R} - m\E[r]\right)}{3}\right)}\right)}\\
        \times\mathbb{P}\bigl(\eta_\text{act}(t) = m \bigr) + \delta_{act}\left(\floor[\bigg]{\frac{\mathcal{R}}{\E\left[r\right]}}\right)\\
        = \gamma^{\dagger}\left(\mathcal{R}\right) \,.
    \end{multline*}
    As a result of Bernstein's inequality, the bound $\gamma^{\dagger}\left({\mathcal{R}}\right)$ is less than $1$ for some interval of $\mathcal{R}\in(\Gamma_a, \infty)$ where it attains the value of $1$ if $\mathcal{R} \leq \Gamma_a$. To find the exact interval for when $\gamma^{\dagger}\left(\mathcal{R}\right) = 1$ requires finding a specific value of $\mathcal{R}$; however, we know that $\Gamma_a$ must be greater than or equal to $\E\left[\eta_\text{act}(t)\right]\E\left[r\right]$ as a result of using Bernstein's inequality on $Q(t)$. Hence, 
    \begin{align*}
        \gamma\left(\mathcal{R}\right) = 
        \begin{cases}
        \min\left\{1, \gamma^{\dagger}\left(\mathcal{R} \right) \right\} & \mathcal{R} > \E\left[\eta_\text{act}(t)\right]\E\left[r\right]\\
        1, & \text{otherwise}.
        \end{cases}
    \end{align*}
    Now, recalling from~\cite[Equation (9)]{massey2002analysis} that $\E\left[\eta_\text{act}(t)\right] = \lambda \E[\theta_{act}]$ and that $\mathbb{P}\left(Q(t)< \mathcal{R}\right) = 1 - \P\bigl(Q(t) \geq  \mathcal{R}\bigr) \geq 1 - \gamma\left(\mathcal{R} \right)$ completes the proof. 
    \end{enumerate}
\end{pf}

Theorem \ref{theorem: mainresult} quantifies the likelihood a charging facility under stochastic user arrivals and charging demand will stay within (or exceed) a specified threshold of user capacity and active user rate consumption. Remarkably, Theorem \ref{theorem: mainresult} is applicable to both the discrete and continuous pricing schemes under their respective assumptions.

\section{Numerical Studies}
\label{section: casestudy}
In this section, we present two numerical studies: a study which illustrates the theoretical results of Theorem \ref{theorem: mainresult} compared to Monte Carlo simulations, and a study which shows how a charging facility operator can utilize the main theorem results to set the charging facility pricing function parameters for both operational models. 

\subsection{Monte Carlo Study of Bounds}
We first study the DSL model and consider a charging facility system which broadcasts $L=4$ pricing functions.\footnote{The code for this case study is available at\\ \url{https://github.com/gtfactslab/automatica_charging_facility}} Satisfying Assumption \ref{assump: rv_arrival}, we suppose uniform distributions for the demand $x$, impatience factor $\alpha$, and time spent at the location $\xi$ with support $[x_\text{min}, x_\text{max}]=[10,100]$ (kWh), $[\alpha_\text{min}, \alpha_\text{max}]=[0,10]$ (\$/hr.), and $[\xi_\text{min}, \xi_\text{max}]=[0,3.5]$ (\$/hr.), respectively. Additionally, the arrival rate for the users to the charging facility is $\lambda = 20$ EVs/hr. The parameters of the random variables for the charging facility pricing function parameters in this case study are in Table \ref{table: chargingfacilitycasestudyparameter}.

\begin{table}[]
    \centering
    \caption{Case Study Parameters for Facility}
    \begin{tabular}{cccl}
    \textbf{Model} & 
    \textbf{Var.} &  \multicolumn{1}{l}{\textbf{Value/Range}} & \textbf{Units}  \\ 
    \hline  
    \textbf{Defined}   & $L$ & 4 & -\\ 
     \textbf{Service Level}   & $R^\ell$ & $15,25,35,45$  & kW\\
                        & $V^\ell$ & $0.20, 0.22, 0.24, 0.26$  & \$/kWh\\
                        &$\mathbb{E}[x/r]$ & 1.87 & hr.  \\
                        &$\mathbb{E}[r]$ & 27.68 & kW \\ 
    \textbf{Prescribed} & $D$ & $2$ & \$/kW-hr\textsuperscript{3} \\
    \textbf{Deadline}   & $B$ & $0.25$ & \$/kWh \\
                        & $\omega$ & 4& hr.\\
                        &$\mathbb{E}[u]$ & 3.92 & hr. \\
                        &$\mathbb{E}[r]$ &  12.60 & kW \\            
    \hline
    \end{tabular}
    \label{table: chargingfacilitycasestudyparameter}
\end{table}

\begin{figure}
        \centering
        \includegraphics[scale = .6]{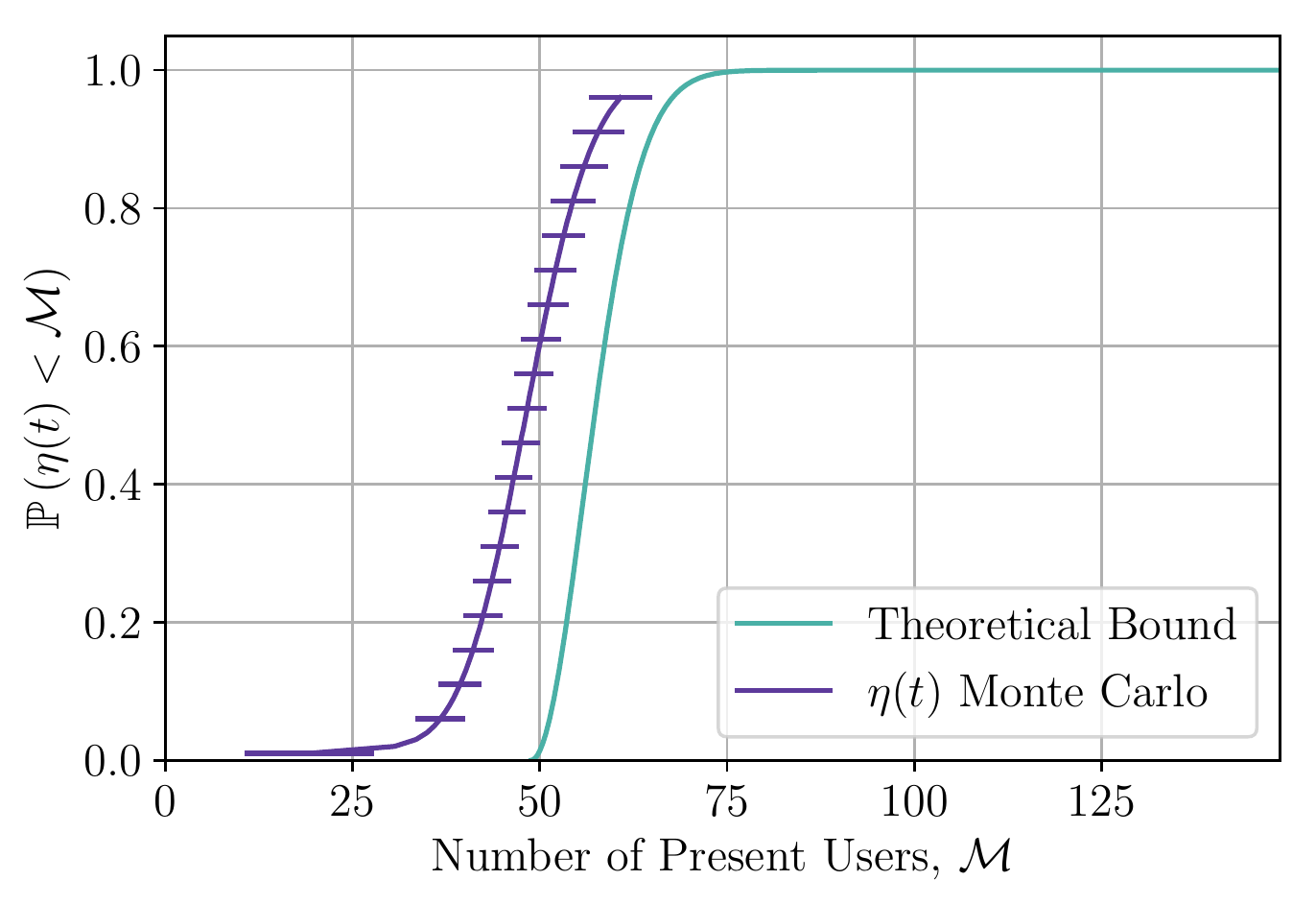}
        \caption{DSL model plot of the theoretical bound, i.e., $1-\delta(\mathcal{M})$, from Theorem~\ref{theorem: mainresult} on the total number of present users in the charging facility versus Monte Carlo results at various percentiles. The error bars represent the values two standard deviations away from the mean in a particular percentile for all Monte Carlo draws while the curve itself connects the mean of each percentile. }
        \label{fig: Nterrorbar}
    \end{figure}

  To illustrate Statement 1 of Theorem~\ref{theorem: mainresult} relating to the number of present users at the charging facility $\eta(t)$, we conduct a $1000$ run Monte Carlo simulation which we  use as a benchmark to illustrate the value of $\P\left(\eta(t) <  \mathcal{M} \right)$. This is shown in Fig.~\ref{fig: Nterrorbar} with error bars that illustrate the values within two standard deviations of the mean attained across all the Monte Carlo runs at specified probability levels. 
  Furthermore, we illustrate the theoretical lower bound on $\P\left(\eta(t) <  \mathcal{M} \right)$, i.e., $1 - \delta(\mathcal{M})$, as a function of $\mathcal{M}$.

  
  Notice that in Fig. 2 when $\mathcal{M} = 55$, we see that the results from Theorem \ref{theorem: mainresult} indicate that $\mathbb{P}\left(\eta(t) < \mathcal{M} \right) \geq 0.35.$ where the Monte Carlo indicates $\mathbb{P}\left(\eta(t) < \mathcal{M} \right)$ is approximately $0.80$. Empirically, we see the theoretical bound on the number of present users in the charging facility for the specified service levels provides operators with the ability to adequately quantify the likelihood $\eta(t)$ will exceed some threshold. \par
\begin{figure}
        \centering
        \includegraphics[scale = .60]{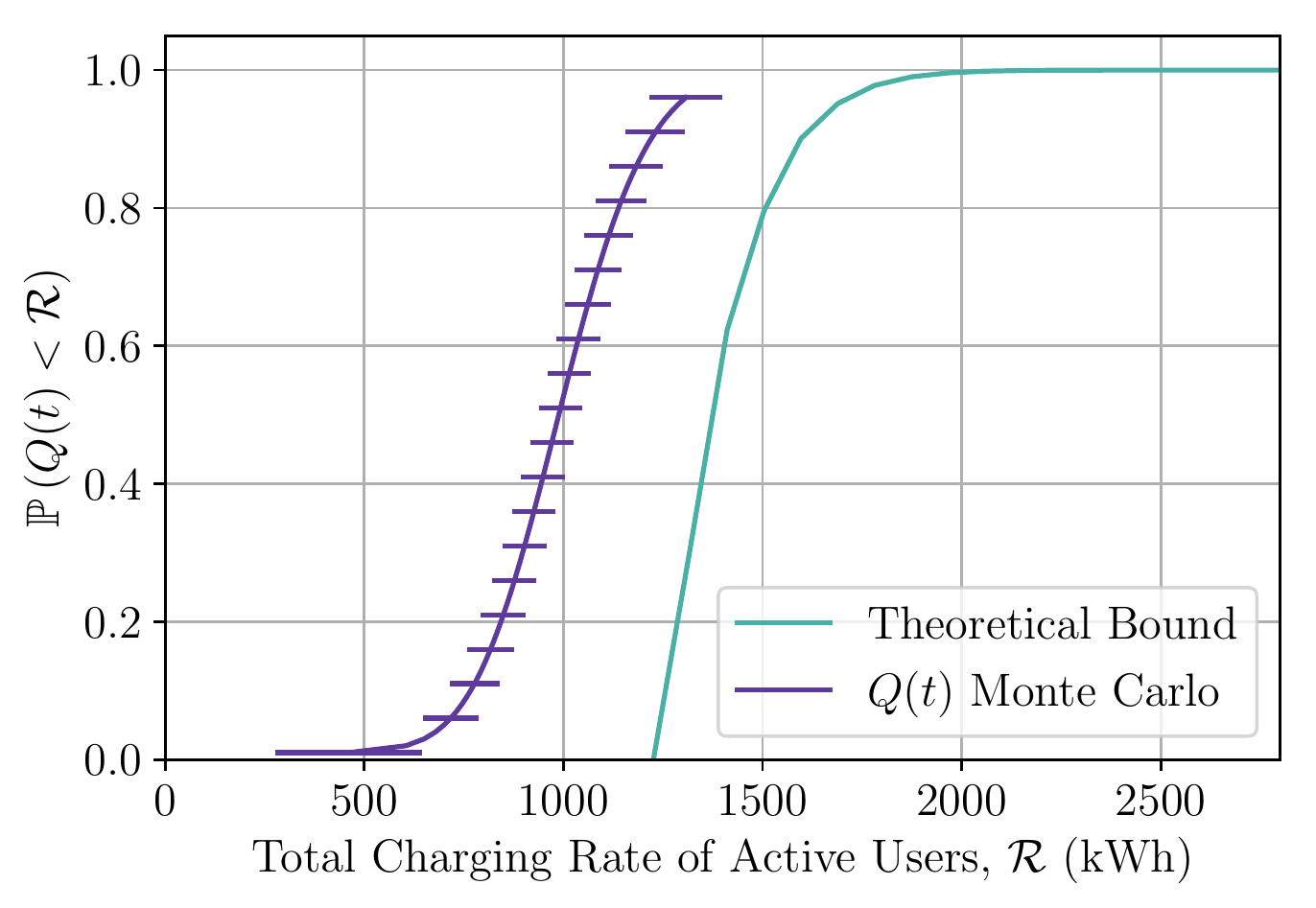}
        \caption{DSL model plot of the theoretical bound, i.e., $1-\gamma(\mathcal{R})$, from Theorem \ref{theorem: mainresult} on the total rate demanded by actively charging users versus Monte Carlo results at various percentiles. The error bars represent the values two standard deviations away from the mean in a particular percentile for all Monte Carlo draws. }
        \label{fig: Qterrorbar}
\end{figure}

Similarly, in this paper, we consider a charging facility that is required to adhere to total power rate constraints from all of its actively charging users. 
For this, we utilize Statement 2 of Theorem \ref{theorem: mainresult}. Fig. \ref{fig: Qterrorbar} shows the theoretical lower bound, i.e., $1 - \gamma(\mathcal{R})$ on the probability the total power draw of the active users will exceed some value~$\mathcal{R}$. The average total power consumption and the two standard deviation error bars across all Monte Carlo runs are presented in Fig. \ref{fig: Qterrorbar}. Here, we see that the theoretical bound provides a conservative quantification of the amount of power draw of the active users in the charging facility. This bound is more conservative than the bound on the total number of active users because it is also dependent on $\delta_\text{act}\left( \mathcal{M}\right)$ from Statement 2 of Theorem \ref{theorem: mainresult} which is itself not an exact account of the number of active users in the charging facility. 

\begin{figure}
    \centering
    \includegraphics[scale=.6]{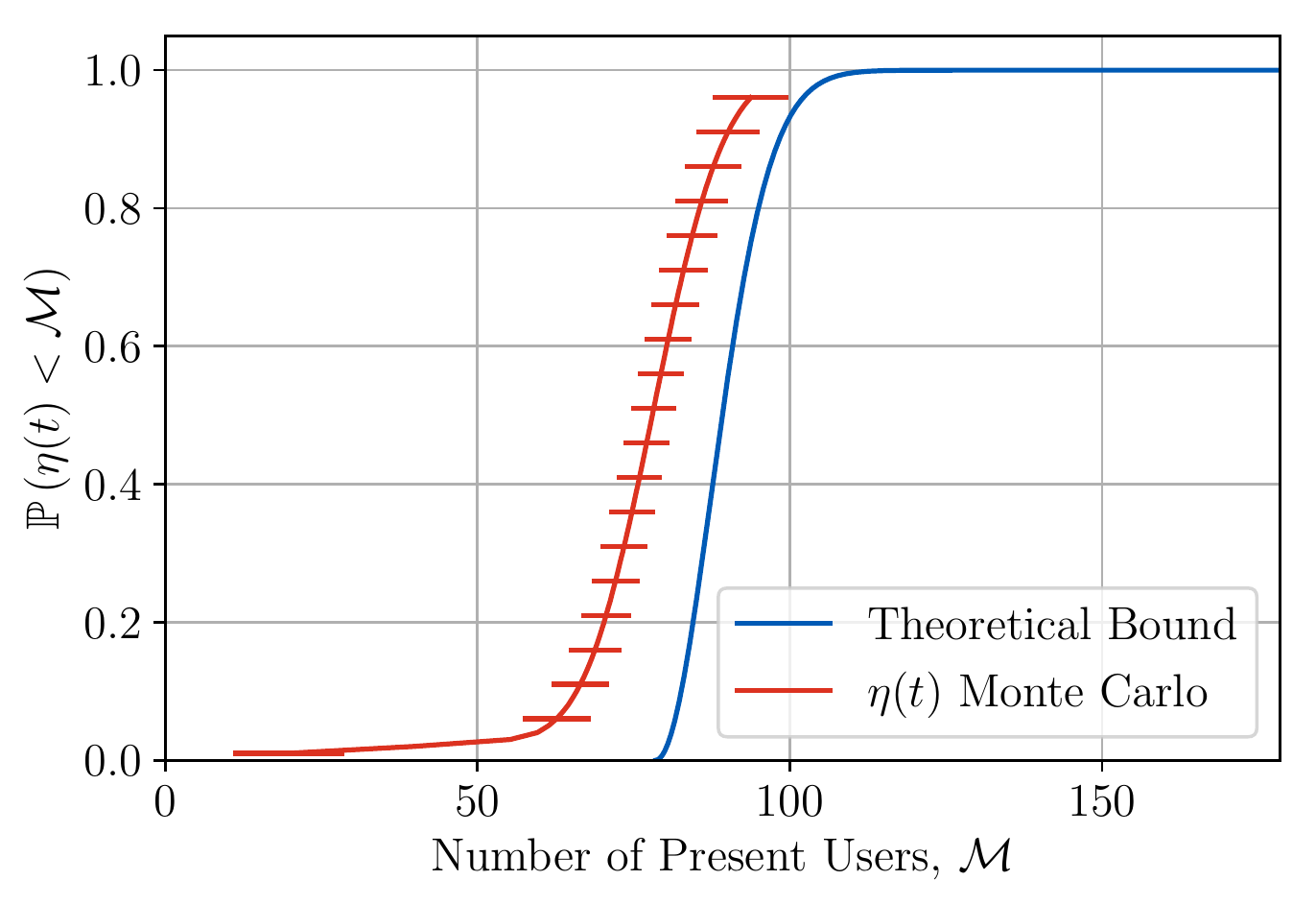}
    \caption{PD model plot of the theoretical bound, i.e., $1-\delta(\mathcal{M})$, from Theorem~\ref{theorem: mainresult} on the total number of present users in the charging facility versus Monte Carlo results at various percentiles. The error bars represent the values two standard deviations away from the mean in a particular percentile for all Monte Carlo draws while the curve itself connects the mean of each percentile.}
    \label{fig: Nterrorbar_continuous}
\end{figure}

\begin{figure}
    \centering
    \includegraphics[scale=.6]{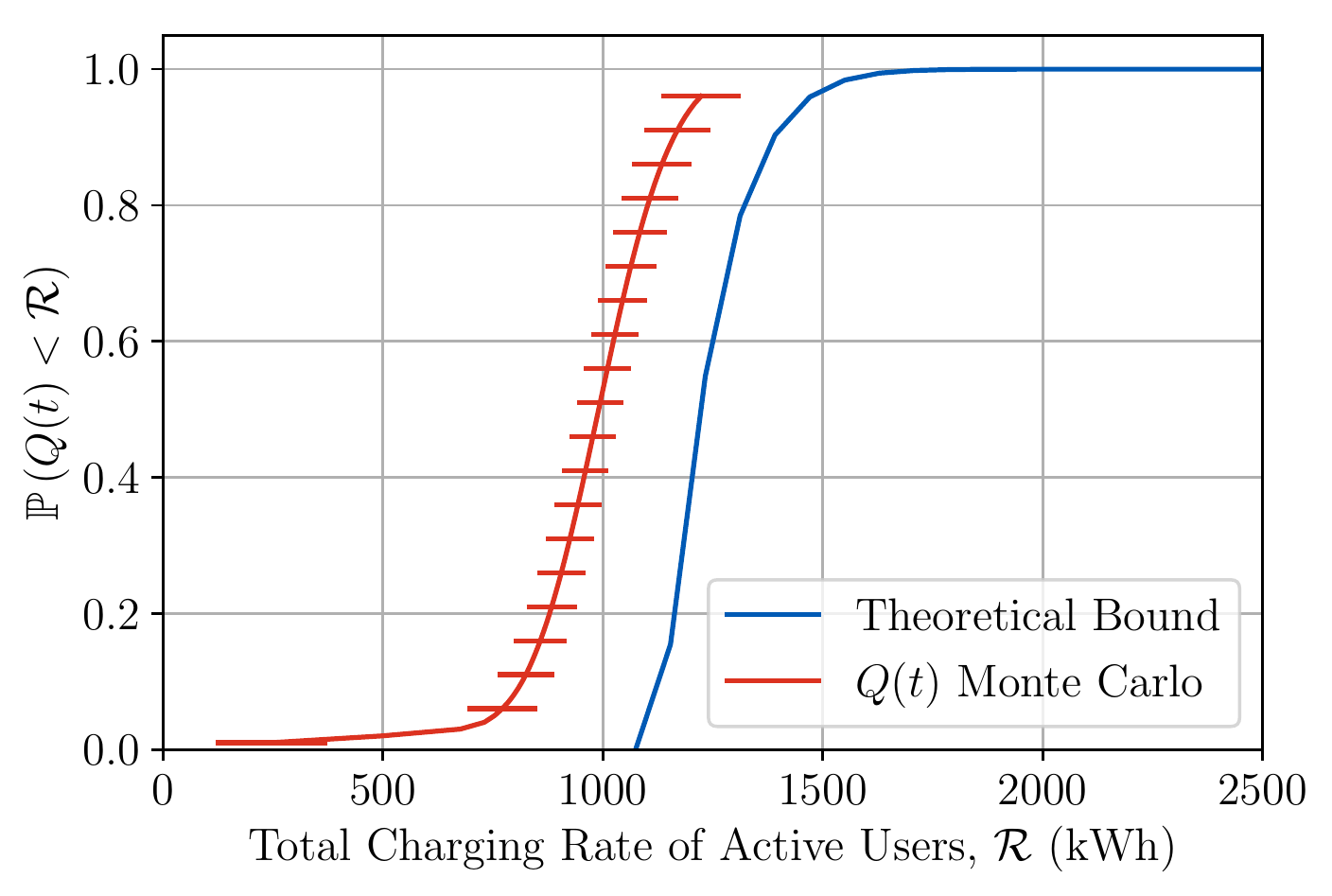}
    \caption{PD model plot of the theoretical bound, i.e., $1-\gamma(\mathcal{R})$, from Theorem \ref{theorem: mainresult} on the total rate demanded by actively charging users versus Monte Carlo results at various percentiles. The error bars represent the values two standard deviations away from the mean in a particular percentile for all Monte Carlo draws.}
    \label{fig: Qterrorbar_continuous}
\end{figure}

Next, we consider the PD model and perform an identical analysis using both statements of Theorem \ref{theorem: mainresult}. We assume the pricing function \eqref{eq: pricingfunc_continuous} from Example \ref{ex:1} and total cost function \eqref{eq: total_cost_func_continuous} with parameters  
listed in Table \ref{table: chargingfacilitycasestudyparameter}. 
Like in the DSL model, we conduct a Monte Carlo simulation and compare the results with the bounds from Theorem~\ref{theorem: mainresult}. These results are shown in in Fig. \ref{fig: Nterrorbar_continuous} and \ref{fig: Qterrorbar_continuous}. 

\subsection{Resource Aware Pricing}
A charging facility operator whose facility operates under the DSL model with total user cost functions as in \eqref{eq: cost_func} and Assumptions \ref{assump: rv_arrival}, \ref{assump: ordering_of_func}, and \ref{assump: timeatcharger} can utilize Theorem \ref{theorem: mainresult} to properly estimate a high-confidence bound on the number of active users using its facilities and their power consumption and subsequently see the effects on $\eta(t)$ and $Q(t)$ resulting from changing the charge rates. To illustrate this point we proceed with a numerical example. 

For the DSL model, computing the high-confidence bounds depends on $\mathbb{E}[x/r]$, $\mathbb{E}[r]$ and $\mathbb{E}[r^2]$. 
As an example, consider an EV charging facility operator with capacity for 40 simultaneously present vehicles and who would like to ensure with high probability that a space is available for each arriving user. 
Therefore, the facility operator would like to quantify the likelihood the number of present users will exceed a specified threshold. Here, an operator can use Statement 1 from Theorem \ref{theorem: mainresult} to get such a bound. 

\begin{figure}
    \centering
    \includegraphics[scale=.50]{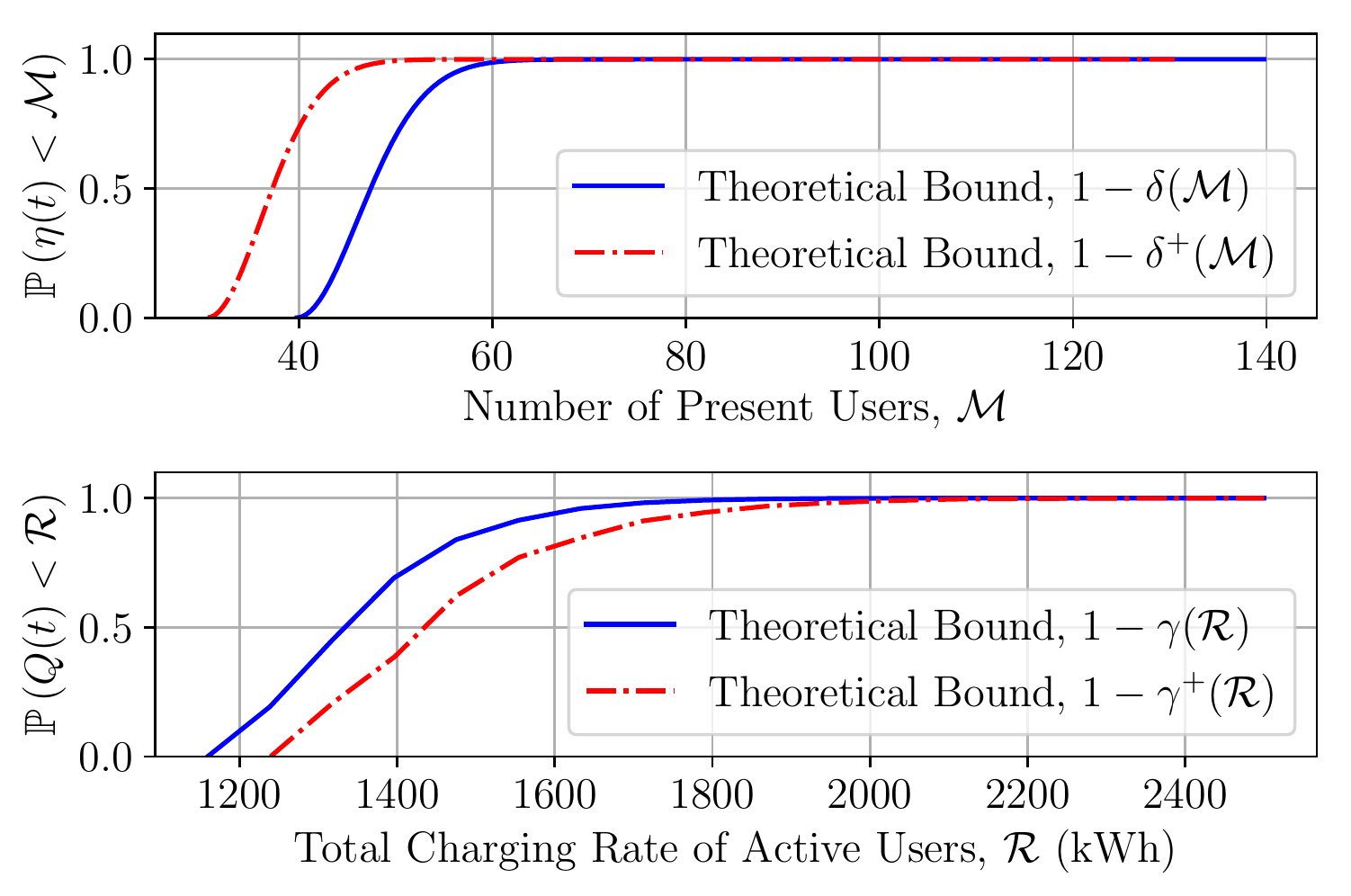}
    \caption{We illustrate the change in the theoretical bounds resulting from an operator increasing the vehicle charge rates in the charging facility in the DSL model. The top plot illustrates the theoretical bound on the number of users, $1 - \delta(\mathcal{M})$, in the charging facility with a baseline charging rate compared to the bound after increasing the charging rates, $1 - \delta^+(\mathcal{M})$. Here, we see that after the charging facility increases the vehicle charging rates there is higher confidence, at a lower level of present users, that the number of present users will be less than $\mathcal{M}$; however, this comes at the cost of a higher total charge rate at the charging facility.}
    \label{fig: illustration_of_theorem}
\end{figure}

For instance, suppose the operator offers $L=2$ service levels with $R^1 = 30$, $R^2 = 40$, $V^1 = 5.2$, and $V^2 = 5.4$. Each arriving user chooses a service level according to~\eqref{eq: level_func}. The resulting theoretical bound on the probability the number of present users is less than $\mathcal{M}$ at the charging facility is illustrated in blue in the upper plot of Fig.~\ref{fig: illustration_of_theorem}.  Notice that the theoretical bound predicts that, for $\mathcal{M} = 40$ active users $\P\left(\eta(t) < \mathcal{M} \right) \geq 0$. This of course is a trivial bound and hence provides little assurance that $\eta(t)$ will not exceed a value of $40$.
\par

If the operator wishes to achieve a higher level of confidence that the facility capacity will not be exceeded, the operator can increase the charging rates to $(R^1)^+ = 50$ and $(R^2)^+ = 70$ while maintaining $V^1$ and $V^2$ the same. Notice that the new theoretical bound, $1 -\delta_{\mathcal{M}}^+(\mathcal{M})$, has increased the confidence that the number of active users will not exceed $40$, i.e., $\P\left(\eta(t) < 40 \right) \geq 0.75$; however, this occurs at the expense of higher total active user charging rates. This is seen in the bottom plot of Fig. \ref{fig: illustration_of_theorem} where $1-\gamma(\mathcal{R})$ shifts to the right to become $1-\gamma^+(\mathcal{R})$ after the charging rates increase, i.e., there is lower confidence the total active user charging rate will not exceed a given total charging rate. Hence, a charging facility operator can use Theorem \ref{theorem: mainresult} to adjust the individual service level charging rates to manage the number of active users at the expense of the charging facility total charging rate. A similar exercise can be conducted for a case when the facility total charging rate is of concern where one would decrease the charging rates. 


In the PD model we demonstrate a similar phenomenon. Consider a charging facility operating with a total cost function as in \eqref{eq: total_cost_func_continuous} under Assumptions \ref{assump: rv_arrival}, \ref{assump: timeatcharger_continuous}, and \ref{assump: chargeratevaluerange}. We can use Theorem \ref{theorem: mainresult} to achieve a desirable confidence bound on the number of present users at the charging facility.

\begin{figure}
    \centering
    \includegraphics[scale=.50]{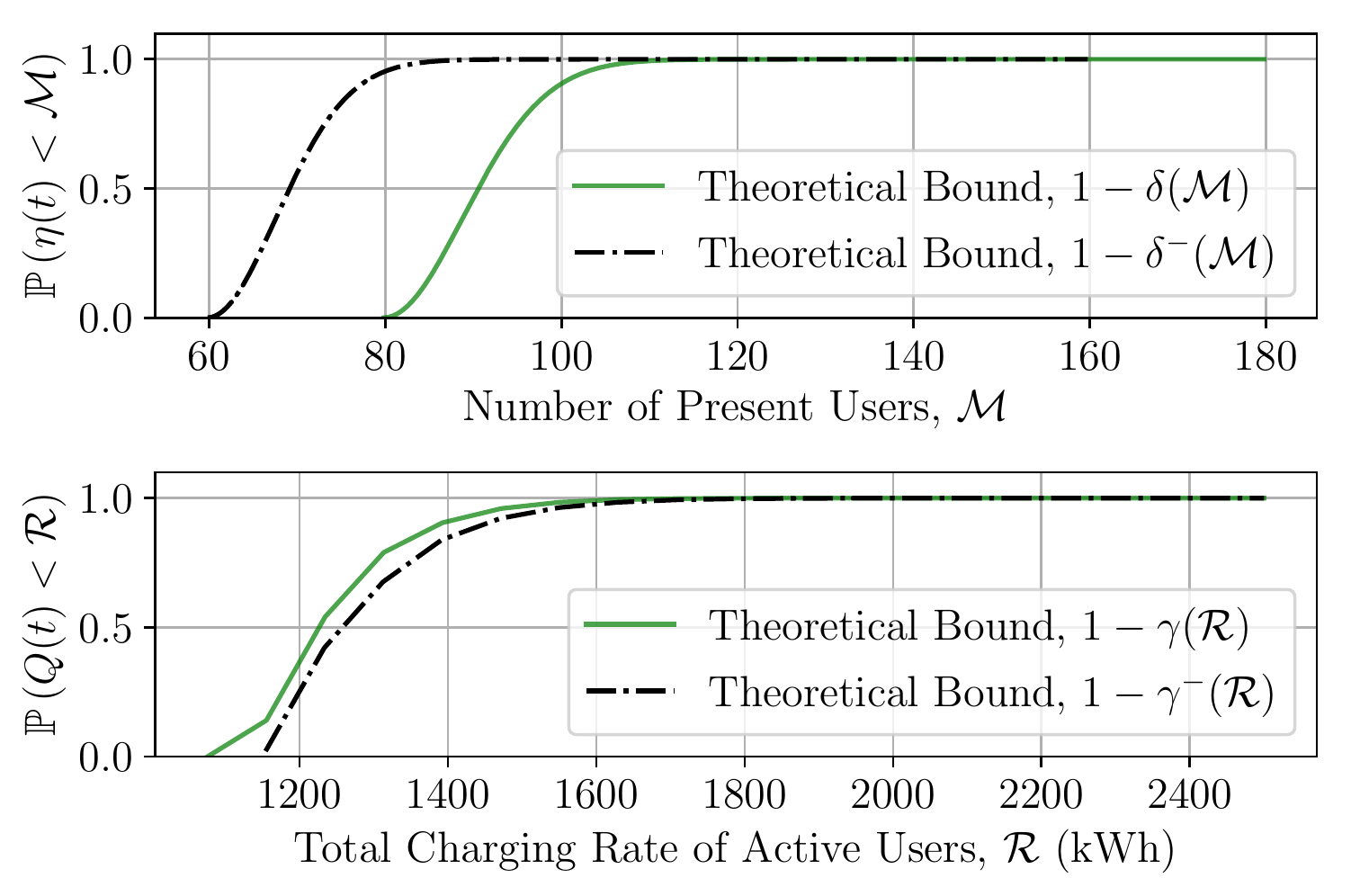}
    \caption{We illustrate the change in the theoretical bounds resulting from an operator decreasing $\omega$ at the charging facility in the PD model. The top plot illustrates the theoretical bound on the number of present users, i.e. $1 - \delta(\mathcal{M})$, in the charging facility with a baseline $\omega$ compared to the bound after decreasing $\omega$, $1 - \delta^-(\mathcal{M})$. Note that after the charging facility decreased $\omega$ there is higher confidence, at lower levels of present users, that the number of present users will be less than $\mathcal{M}$; however, this comes at the cost of a higher total charge rate at the charging facility.}
    \label{fig: illustration_of_theorem_continuous}
\end{figure}

For instance, the operator offers a pricing function as in~\eqref{eq: pricingfunc_continuous} where $D = 2$, $B = 5$, $\omega = 4$, and $R^{\text{max}} = 50$ and where the total cost that users are trying to minimize is~\eqref{eq: total_cost_func_continuous}. Note that in this model the charging operator deals with deadlines and hence they will adjust the parameter $\omega$ accordingly. The resulting theoretical bound on the number of active users for various confidence bounds is illustrated in green in the upper plot of Fig. \ref{fig: illustration_of_theorem_continuous}.  Notice that the theoretical bound predicts that, for $\mathcal{M} = 80$ active users $\P\left(\eta(t) < 80 \right) \geq 0$. This is a trivial bound and hence provides little confidence that $\eta(t)$ will not exceed a value of $80$.

Hence, to achieve a higher level of confidence that $\eta(t)$ will be less than $80$, a charging operator can adjust $\omega$ to $(\omega)^- = 2.5$. Notice in Fig. \ref{fig: illustration_of_theorem_continuous} that the new theoretical bound, $1 -\delta_{\mathcal{M}}^-(\mathcal{M})$, has increased the confidence that the number of active users will not exceed $80$, i.e., $\P\left(\eta(t) < 80 \right) \geq .95$; however, this occurs at a slight expense of higher total active user charging rates. A similar exercise can be conducted for a case when the facility total charging rate is of concern where a charging facility operator would increase~$\omega$.

\section{Conclusion}
\label{section: conclusion}
We study the problem of providing probabilistic bounds on an EV charging facility's likelihood of exceeding a specified number of present users and the total active user power draw. Specifically, we focus on charging facilities which can deploy either a defined service level (DSL) model, i.e., where users choose from finitely many charging rates, or a prescribed deadline (PD) model, i.e., where users choose a charging deadline. In both models, we leverage knowledge on the probability distributions of the user parameters to provide probabilistic guarantees. We illustrate these derived probabilistic bounds in a case study and ultimately demonstrate how a charging facility operator can utilize these results to set charging facility parameters in order to achieve desired behavior.

\bibliographystyle{plain} 
\bibliography{references.bib}

\appendix
\section{Proofs and Additional Propositions}
\subsection{Proof of Proposition \ref{proposition: probability_choose_level}}\label{app: propmin}
\vspace{-1.5em}
\begin{pf}
    Under Assumptions \ref{assump: rv_arrival}, \ref{assump: ordering_of_func}, and \ref{assump: timeatcharger}, consider the set of $L$ pricing functions 
    $\big\{g_\ell(x_j,\alpha_j, \xi_j) \big\}^L_{\ell = 1}$ of the form $ g_\ell(x_j, \alpha_j, \xi_j) = x_jV^\ell  + \alpha_j\left[ \frac{x_j}{R^\ell} - \xi_j\right]_+ + F \left[\xi_j - \frac{x_j}{R^\ell} \right]_+$ 
    as specified in \eqref{eq: cost_func}. 
    From \eqref{eq: level_func}, and the law of total probability, we  have that
    \begin{multline*}
         \P(S(x_j, \alpha_j, \xi_j) = k) = \int_{\rho_j} \P\bigl( g_k = \min_{i} g_i\mid \rho_j \bigr) f_{\Rho}(\rho_j) \d\Rho \\ = \E_\Rho\left[\P\left( g_k = \min_{i} g_i\mid \rho_j \right)\right]\,,
    \end{multline*}
    where we recall the random variable $\rho_j = x_j/\xi_j$, and, for convenience, we sometimes omit the arguments of the pricing functions. In the remainder of the proof, we establish closed form expressions for $\P\bigl( g_k = \min_{i} g_i\mid \rho_j \bigr)$ by considering the cases corresponding to the interval partitions introduced in Section \ref{subsec: discretepricing_probabilityoflevel}, namely, the intervals 
    $\rho_j < R^1$, $\rho_j \in [R^m, R^{m+1})$ for $m \in [1, \dots ,L-1]$, and $R^L < \rho_j$. 
    
    For future use, define $\vdiff{k}{i}{V} = V^k - V^i$ and $\vdiff{k}{i}{\bar{R}}=\bar{R}^k-\bar{R}^i=1/R^i-1/R^k$ for all $i, k$. Throughout the proof, we will use the observation that $g_k(x_j,\alpha_j, \xi_j) = \min_{i} g_i(x_j,\alpha_j, \xi_j)$ 
    if and only if $g_k(x_j,\alpha_j, \xi_j) \leq g_i(x_j,\alpha_j, \xi_j)$ for all $i$.

    \textbf{\underline{Case 1: $\mathbf{\boldsymbol{\rho}_j < R^1 \, \,\,}$}} \\
    When $\rho_j < R^1$ this implies that $\rho_j$ is less than all charging rates as a result of the ordering of the service levels, and as a result with $k = 1$ we get $g_1(x_j, \alpha_j, \xi_j) - g_i(x_j, \alpha_j, \xi_j) = x_j(\vdiff{1}{i}{V} - F \vdiff{1}{i}{\bar{R}})$.
    Note that due to the ordering delineated in Assumption~\ref{assump: ordering_of_func}, this quantity is always less than zero since $\vdiff{1}{i}{V} < 0$, $\vdiff{1}{i}{\bar{R}} > 0$, $F >0$, and $x_j > 0$. For any other choice of $k \in \left\{ 2, \dots, L\right\}$, there exists $i \neq k$ such that $g_k(x_j, \alpha_j, \xi_j) - g_i(x_j, \alpha_j, \xi_j) = x_j (\vdiff{k}{i}{V} - F \vdiff{k}{i}{\bar{R}}) > 0$, and hence such a choice of $k$ can not be the minimum.
    Hence, we obtain the conditional probability in the case when $\rho_j < R^1$ that $\P( g_k = \min_{i} g_i\mid \rho_j ) = 1$  if $k = 1$ and  $\P( g_k = \min_{i} g_i\mid \rho_j ) = 0$ if $k \neq 1$. 
    
    \textbf{\underline{Case 2: $\mathbf{\boldsymbol{\rho_j} \boldsymbol{\in} [R^m, R^{m+1})}$}}\\ 
    First, consider the case when the minimizing index $k \leq m$ for some $m\in \{1, \dots, L-1\}$. Then 
    \begin{multline*}
    g_k(x_j, \alpha_j, \xi_j) - g_i(x_j, \alpha_j, \xi_j) =  \\ x_j \vdiff{k}{i}{V} + \alpha_j  \left( \frac{x_j}{R^k} - \xi_j - \left[   \frac{x_j}{R^i} - \xi_j\right]_+ \right) - F \left( \left[ \xi_j - \frac{x_j}{R^i} \right]_+  \right) \,.
    \end{multline*}
    Hence, we consider several cases for $i$. When $i < k$, $g_k(x_j, \alpha_j, \xi_j) - g_i(x_j, \alpha_j, \xi_j) =  x_j(\vdiff{k}{i}{V} + \alpha_j \vdiff{k}{i}{\bar R})$. Notice that, since $x_j>0$, the sign of this difference depends only on the random variable $\alpha_j$. Since $k$ is assumed to be the minimizing index, this difference must be nonpositive for all $i$. 
    Rearranging, we see that $g_k$  being the minimizer for some $k\leq m$ implies  $\alpha_j > {(\vdiff{i}{k}{V})}/{(\vdiff{k}{i}{\bar R})}$ for all $i<k\leq m$.
    
    Similarly, for $k < i \leq m$, the difference $g_k(x_j, \alpha_j, \xi_j) - g_i(x_j, \alpha_j, \xi_j) = x_j(\vdiff{k}{i}{V} + \alpha_j \vdiff{k}{i}{\bar R} )$.
    This difference is negative when $\alpha_j < {(\vdiff{i}{k}{V})}/{(\vdiff{k}{i}{\bar R})}$.
    Lastly, when $m+1 \leq i$, 
    \begin{multline*}
        g_k(x_j, \alpha_j, \xi_j) - g_i(x_j, \alpha_j, \xi_j) = \\ x_j\vdiff{k}{i}{V} + \alpha_j \left(  \frac{x_j}{R^k} - \xi_j \right) - F \left(\xi_j - \frac{x_j}{R^i}\right)\,.
    \end{multline*}
    Similarly as before, after algebraic manipulation this quantity is negative when
    \begin{align*}
        \alpha_j < \frac{F \left(\frac{1}{\rho_j} - \frac{1}{R^i} \right) - \vdiff{k}{i}{V}}{\frac{1}{R^k} - \frac{1}{\rho_j} } \,.
    \end{align*}
    Combining the above inequalities on $\alpha_j$, and defining $f_A(\alpha_j)$ to be the probability distribution of $\alpha_j$, this establishes that, 
     when $\rho_j \in [R^m, R^{m+1})$ and $k \leq m$, 
    \begin{multline*}
        \P\biggl( g_k = \min_{i} g_i\mid \rho_j \biggr) = \left[ \int^{\Bar{\alpha}^k_1}_{\ubar{\alpha}^k_1} f_A(\alpha_j) \d A \right]_+ \\= \P\left(\ubar{\alpha}^k_1 < \alpha_j < \Bar{\alpha}^k_1 \right)\,,
    \end{multline*}
    where $\Bar{\alpha}^k_1$ and $\ubar{\alpha}^k_1$ are as defined in the statement of Proposition \ref{proposition: probability_choose_level}.
    Now consider the case when $k = m+1$. Then 
    \begin{multline*}
        g_k(x_j, \alpha_j, \xi_j) - g_i(x_j, \alpha_j, \xi_j) = \\ x_j\vdiff{k}{i}{V}  - \alpha_j \left[ \frac{x_j}{R^i} - \xi_j\right]_+   + F \left(  \xi_j -  \frac{x_j}{R^k} -  \left[ \xi_j - \frac{x_j}{R^i} \right]_+  \right) \,.
    \end{multline*}
    for all $i$. 
    Consider first the case when $i < m + 1$. Then the difference becomes 
    \begin{multline*}
        g_k(x_j, \alpha_j, \xi_j) - g_i(x_j, \alpha_j, \xi_j) = \\ x_j\vdiff{k}{i}{V}  - \alpha_j \left(\frac{x_j}{R^i} - \xi_j \right) + F \left(  \xi_j -  \frac{x_j}{R^k} \right)\,,
    \end{multline*}
    which is negative when
    \begin{align*}
        \alpha_j > \frac{F \left(\frac{1}{R^k} - \frac{1}{\rho_j} \right) - \vdiff{k}{i}{V}}{\frac{1}{\rho_j} - \frac{1}{R^i}}\,.
    \end{align*}
    Similarly, still considering the case where $k = m+1$, when $m + 1 < i$,  the difference becomes
    $ g_k(x_j, \alpha_j, \xi_j) - g_i(x_j, \alpha_j, \xi_j) = x_j \left( \vdiff{k}{i}{V} - F \vdiff{k}{i}{\bar R} \right)$, which is always negative when $i > m+1$. Hence,  when $\rho_j \in [R^m, R^{m+1})$ and $k = m + 1$, the quantity
    \begin{multline*}
        \P\left( g_k = \min_{i} g_i\mid \rho_j \right) = \left[ \int^{\alpha_\text{max}}_{\ubar{\alpha}^k_2} f_A(\alpha_j) \d A \right]_+ \\ =\P\left(\ubar{\alpha}^k_2 < \alpha_j < \alpha_\text{max} \right)\,, 
    \end{multline*}
    where $\ubar{\alpha}^k_2$ is as defined in the statement of Proposition \ref{proposition: probability_choose_level}.
    
    Lastly, consider the case when $ k > m+1$. For some $i\geq m+1$, $ g_k(x_j, \alpha_j, \xi_j) - g_i(x_j, \alpha_j, \xi_j)  = \vdiff{k}{i}{V} - F \vdiff{k}{i}{\bar R}>0$, and thus $k$ cannot be the minimizing index. As a result, we have that if $\rho_j \in [R^m, R^{m+1})$ and $k>m+1$, $\P(g_k = \min_i g_i \mid \rho_j ) = 0$.

    
    \textbf{\underline{Case 3: $\mathbf{R^L < \boldsymbol{\rho}_j }$}}
    \\
    When $R^L < \rho_j$ this implies that $\rho_j$ is greater than all charging rates as a result of the ordering of the service levels. Moreover, for all $k$ and $i$, 
    $g_k(x_j, \alpha_j, \xi_j) - g_i(x_j, \alpha_j, \xi_j) = x_j( \vdiff{k}{i}{V} + \alpha_j \vdiff{k}{i}{\bar R})$. Again, since $x_j>0$, the sign of this difference depends only on the random variable $\alpha_j$. 
    In particular, the difference is negative when $\alpha_j < {(\vdiff{i}{k}{V})}/{(\vdiff{k}{i}{\bar R})}$. 
    Combining these inequalities for all $i$, It follows that when $R^L < \rho_j$, 
    \begin{multline*}
        \P(g_k = \min_i g_i \mid \rho_j) = \left[ \int^{\bar{\alpha}^k_3}_{\ubar{\alpha}^k_3} f_A\left\{\alpha_j \right\}\d A \right]_+ \\ =\P\left(\ubar{\alpha}^k_3 < \alpha_j < \Bar{\alpha}^k_3 \right) \,,
    \end{multline*}
    where $\Bar{\alpha}^k_3$ and $\ubar{\alpha}^k_3$ are as defined in the statement of Proposition \ref{proposition: probability_choose_level}.
    This completes the proof.
\end{pf}

\subsection{Proposition~\ref{prop: BernoullitoRV}}
\label{sec:appendix_obs}
\begin{prop}
\label{prop: BernoullitoRV}
Let $Z$ be a Poisson random variable with mean $\bar\lambda$. Then, for any $\mathcal{M} > \bar\lambda > 0$, it holds that $\P\bigl(Z < \mathcal{M} \bigr) \geq 1 - \delta(\mathcal{M})$, where 
\begin{align*} 
    \delta(\mathcal{M}) = \exp{\left(\frac{-\left(\mathcal{M} - \bar\lambda \right)^2}{2\left(\bar\lambda + \frac{\mathcal{M} - \bar\lambda}{3}\right)}\right)} \,.
\end{align*}
\end{prop}
Before proving Proposition \ref{prop: BernoullitoRV}, we recall Bernstein's inequality which gives a probabilistic upper bound on the sum of the deviation from the mean of a bounded random variable which is the basis for the proof of the proposition.  
\begin{fact}[Bernstein's Inequality, \cite{wainwright2019highdimstats}]
    \label{defintion-fact: BernsteinsIneq}
    Given $n$ independent, zero-mean random variables $X_i$ such that, for some $b > 0$, $\nu > 0$, $0 \leq X_i \leq b$  almost surely for all $1\leq i\leq n$. Then, it holds that
    \begin{multline}
        \label{eq: bernsteinsdefinition}
        \P\left(\sum^n_{i=1} (X_i - \E[X_i] \big) \geq \nu  \right)\\
        \leq \exp{\left(\frac{-\nu^2}{2\left(\sum^n_{i=1}\E[X_i^2] + \frac{b \nu}{3}\right)}\right)}\,. 
    \end{multline}
\end{fact}
We will now apply the Fact~\ref{defintion-fact: BernsteinsIneq} to prove Proposition \ref{prop: BernoullitoRV}.
\begin{pf}[Proof of Proposition \ref{prop: BernoullitoRV}.]
    We seek to prove a bound on the likelihood a Poisson RV will exceed some value $\mathcal{M}$. Recall from the Poisson limit theorem \cite[Theorem 3.6.1]{Durrett2019} that a Poisson RV $Z$ with mean $\bar\lambda$ can be seen as a sum of $n$ Bernoulli RVs $X_i\leq 1 $ with mean $p$, where $p$ is such that $np \rightarrow \bar\lambda$ when $n \rightarrow +\infty$. In other words, $\sum_{i=1}^n X_i \rightarrow Z$ as $n \rightarrow +\infty$. Here, we see that we can now apply Fact \ref{defintion-fact: BernsteinsIneq} to find a bound on the value of a Poisson random variable which is approximated as the sum of Bernoulli RVs. 

    Let $X = \sum_{i=1}^n X_i$ and $\E[X] = \E\left[ \sum_{i=1}^n X_i \right]= \sum_{i=1}^n \E\left[X_i\right] = np$. Since Fact \ref{defintion-fact: BernsteinsIneq} applies to zero-mean random variables, let $X^0 = X - \E[X]= \sum_{i=1}^n X_i - \sum_{i=1}^n \E\left[X_i\right]$ be a zero-mean sum of Bernoulli random variables where $\E[X^0] = 0$. Then, applying Fact \ref{defintion-fact: BernsteinsIneq} with $b = 1$ and letting $\mathcal{M} = \nu + \E\left[X\right] = \nu + np$,
    \begin{multline*}
        \label{eq: bernsteinsdefinition}
        \P\left(X^0 \geq \nu  \right)
        =\P\left( \sum_{i=1}^n X_i - \sum_{i=1}^n \E\left[X_i\right] \geq \nu   \right) \\
        = \P\left(X - \E[X] \geq \nu \right) 
        = \P\left(X  \geq  \nu + \E[X] \right) \\ 
        = \P\left(X  \geq \mathcal{M} \right)
        \leq \exp{\left(\frac{-\left(\mathcal{M} - np \right)^2}{2\left(np + \frac{\mathcal{M} - np }{3}\right)}\right)}\,. 
    \end{multline*}
    The last inequality uses the fact that $\sum_{i=1}^n \E\left[X_i^2\right] = np$. Since we can approximate a Poisson random variable $Z$ via the Poisson limit theorem, by  letting $n \rightarrow +\infty$, we get
    \begin{align*}
        \P\left(Z  \geq \mathcal{M} \right) \leq \exp{\left(\frac{-\left(\mathcal{M} - \bar \lambda \right)^2}{2\left(\bar\lambda  + \frac{\mathcal{M} - \bar\lambda }{3}\right)}\right)} \,.
    \end{align*}
    This proves the proposition. 
\end{pf}

\end{document}

%% file: plots/fig1.tikz
\begin{tikzpicture}
\begin{axis}[
stack plots=y, 
area style,
enlarge x limits=false, 
enlarge y limits=false, 
ymin=0, 
xmin=10,
title style={yshift=1.5ex},
height=5cm, width=8cm,
legend style={at={(0.5 ,-0.3)},anchor=north}, 
legend columns=2,
title={$\mathbb{P}\left( g_k = \min_{i} g_i\mid \rho_j \right)$}, 
xlabel={$\rho_j$},
ylabel={Probability},
clip=false]

\draw[dotted, thick] (axis cs: 15,0) -- (axis cs: 15,1) node[above] {$R^1$};
\draw[dotted, thick] (axis cs: 25,0) -- (axis cs: 25,1) node[above] {$R^2$};
\draw[dotted, thick] (axis cs: 35,0) -- (axis cs: 35,1) node[above] {$R^3$};
\draw[dotted, thick] (axis cs: 45,0) -- (axis cs: 45,1) node[above] {$R^4$};

\addplot[smooth, mark=none, fill=mycolor1] table[x index=0, y index =1, col sep=space]{plots/fig1data.txt} \closedcycle;
\addplot[smooth, mark=none, fill=mycolor2] table[x index=0, y index =2, col sep=space]{plots/fig1data.txt}  \closedcycle;
\addplot[smooth, mark=none, fill=mycolor3] table[x index=0, y index =3, col sep=space]{plots/fig1data.txt}  \closedcycle;
\addplot[smooth, mark=none, fill=mycolor4] table[x index=0, y index =4, col sep=space]{plots/fig1data.txt}  \closedcycle;
\legend{$k=1$,$k=2$,$k=3$, $k=4$};
\end{axis}
\end{tikzpicture}

%% file: main.bbl
\begin{thebibliography}{10}

\bibitem{alizadeh2018retail}
Mahnoosh Alizadeh, Hoi-To Wai, Andrea Goldsmith, and Anna Scaglione.
\newblock Retail and wholesale electricity pricing considering electric vehicle
  mobility.
\newblock {\em IEEE Transactions on Control of Network Systems}, 6(1):249--260,
  2018.

\bibitem{bae2011spatial}
Sungwoo Bae and Alexis Kwasinski.
\newblock Spatial and temporal model of electric vehicle charging demand.
\newblock {\em IEEE Transactions on Smart Grid}, 3(1):394--403, 2011.

\bibitem{Durrett2019}
Rick Durrett.
\newblock {\em Probability: Theory and Examples}.
\newblock Cambridge University Press, New York, NY, USA, 5th edition, 2019.

\bibitem{gan2012optimal}
Lingwen Gan, Ufuk Topcu, and Steven~H Low.
\newblock Optimal decentralized protocol for electric vehicle charging.
\newblock {\em IEEE Transactions on Power Systems}, 28(2):940--951, 2012.

\bibitem{nrel}
National Renewable~Energy Laboratory.
\newblock {\em Electric Vehicle Pricing and Payments}.
\newblock https://www.nrel.gov/about/ev-charging-stations-pricing.html.

\bibitem{le2016optimal}
Caroline Le~Floch, Francois Belletti, and Scott Moura.
\newblock Optimal charging of electric vehicles for load shaping: A
  dual-splitting framework with explicit convergence bounds.
\newblock {\em IEEE Transactions on Transportation Electrification},
  2(2):190--199, 2016.

\bibitem{li2013distribution}
Ruoyang Li, Qiuwei Wu, and Shmuel~S Oren.
\newblock Distribution locational marginal pricing for optimal electric vehicle
  charging management.
\newblock {\em IEEE Transactions on Power Systems}, 29(1):203--211, 2013.

\bibitem{massey2002analysis}
William~A Massey.
\newblock The analysis of queues with time-varying rates for telecommunication
  models.
\newblock {\em Telecommunication Systems}, 21(2-4):173--204, 2002.

\bibitem{bloombergnef}
Colin McKerracher, Ali Izadi-Najafabadi, Aleksandra O'Donovan, Nick Albanese,
  Nikolas Soulopolous, David Doherty, Milo Boers, Ryan Fisher, Corey Cantor,
  James Frith, Siyi Mi, and Andrew Grant.
\newblock Electric vehicle outlook 2020.
\newblock {\em Bloomberg New Energy Finance}, 2020.

\bibitem{moradipari2019pricing}
Ahmadreza Moradipari and Mahnoosh Alizadeh.
\newblock Pricing and routing mechanisms for differentiated services in an
  electric vehicle public charging station network.
\newblock {\em arXiv preprint arXiv:1903.06388}, 2019.

\bibitem{energygov}
Office of~Energy~Efficiency and Renewable Energy.
\newblock {\em Vehicle Charging}.
\newblock https://www.energy.gov/eere/electricvehicles/vehicle-charging.

\bibitem{ortega2012electric}
Miguel~A Ortega-Vazquez, Francois Bouffard, and Vera Silva.
\newblock Electric vehicle aggregator/system operator coordination for charging
  scheduling and services procurement.
\newblock {\em IEEE Transactions on Power Systems}, 28(2):1806--1815, 2012.

\bibitem{pandit2018discount}
Parthe Pandit and Samuel Coogan.
\newblock Discount-based pricing and capacity planning for ev charging under
  stochastic demand.
\newblock In {\em 2018 Annual American Control Conference (ACC)}, pages
  6273--6278. IEEE, 2018.

\bibitem{santoyo2020multi}
Cesar Santoyo, Gustav Nilsson, and Samuel Coogan.
\newblock Multi-level electric vehicle charging facilities with limited
  resources.
\newblock In {\em IFAC World Congress}, 2020.

\bibitem{sassi2017electric}
Ons Sassi and Ammar Oulamara.
\newblock Electric vehicle scheduling and optimal charging problem: complexity,
  exact and heuristic approaches.
\newblock {\em International Journal of Production Research}, 55(2):519--535,
  2017.

\bibitem{wainwright2019highdimstats}
Martin~J Wainwright.
\newblock {\em High-dimensional statistics: A non-asymptotic viewpoint},
  volume~48.
\newblock Cambridge University Press, 2019.

\bibitem{wu2011vehicle}
Chenye Wu, Hamed Mohsenian-Rad, and Jianwei Huang.
\newblock Vehicle-to-aggregator interaction game.
\newblock {\em IEEE Transactions on Smart Grid}, 3(1):434--442, 2011.

\bibitem{zhang2015game}
Lei Zhang and Yaoyu Li.
\newblock A game-theoretic approach to optimal scheduling of parking-lot
  electric vehicle charging.
\newblock {\em IEEE Transactions on Vehicular Technology}, 65(6):4068--4078,
  2015.

\end{thebibliography}
